\documentclass{ws-ijmpe}

\usepackage{bbm}
\usepackage{graphicx}
\usepackage{amsmath}
\usepackage{amsfonts,amsbsy}
\usepackage{amssymb}
\usepackage{mciteplus}

\newcommand{\ud}{\, \mathrm{d}}

\newcommand{\nc}{{N_\mathrm{c}}}

\newcommand{\half}{\frac{1}{2}}

\newcommand{\lqcd}{\Lambda_{\mathrm{QCD}}}
\newcommand{\as}{\alpha_{\mathrm{s}}}
\newcommand{\Tr}{\mathrm{Tr}}

\newcommand{\cf}{C_\mathrm{F}}

\newcommand{\nr}[1]{(\ref{#1})} 

\newcommand{\ra}{R_A}
\newcommand{\rp}{R_p}
\newcommand{\rpa}{R_{pA}}

\newcommand{\gev}{\ \textrm{GeV}}

\newcommand{\qs}{{Q_\mathrm{s}}}
\newcommand{\qsa}{{Q^A_\mathrm{s}}}
\newcommand{\qsb}{{Q^B_\mathrm{s}}}

\newcommand{\fig}{Fig.~}

\newcommand{\eq}{Eq.~}
\newcommand{\se}{Sec.~}
\newcommand{\eqs}{Eqs.~}

\newcommand{\npart}{{N_\textrm{part}}}
\newcommand{\ncoll}{{N_\textrm{coll}}}
\newcommand{\nparta}{{n_\textrm{part,A}}}
\newcommand{\npartb}{{n_\textrm{part,B}}}

\newcommand{\ncs}{N_\textrm{CS}}

\newcommand{\rt}{\mathbf{r}_T}
\newcommand{\ut}{\mathbf{u}_T}

\newcommand{\bt}{\mathbf{b}_T}
\newcommand{\xt}{\mathbf{x}_T}
\newcommand{\yt}{\mathbf{y}_T}

\newcommand{\pt}{{\mathbf{p}_T}}
\newcommand{\ptt}{p_T} 
\newcommand{\ktt}{k_T} 
\newcommand{\rtt}{r_T} 
\newcommand{\qt}{\mathbf{q}_T}
\newcommand{\kt}{\mathbf{k}_T}

\newcommand{\nabt}{\boldsymbol{\nabla}_T}

\newcommand{\p}{{\mathbf{p}}}
\newcommand{\q}{{\mathbf{q}}}

\begin{document}

\catchline{}{}{}{}{}

\title{
Small $x$ physics and RHIC data
}

\author{\footnotesize T. Lappi}

\address{
Physics Department, P.O. Box 35,
 40014, University of Jyv\"{a}skyl\"{a}, Finland and \\
Helsinki Institute of Physics, P.O. Box 64, 00014 University of Helsinki, Finland
\\
tuomas.lappi@jyu.fi}

\maketitle

\begin{history}
\end{history}

\begin{abstract}
This is a  review of applications of the Color Glass Condensate to the 
phenomenology of relativistic heavy ion collisions. 
The initial stages of the collision can be understood in terms
of the nonperturbatively strong nonlinear glasma color fields.
We discuss how the CGC framework 
can and has been used to compute properties of the initial 
conditions of AA collisions. 
In particular this has led to recent progress in understanding multiparticle 
correlations, which can provide  a directly observable signal of
the properties of the initial stage of the collision process.
\end{abstract}

\newpage

\tableofcontents
\markboth{T. Lappi}{Small $x$ physics and RHIC data}

\newpage

\section{Introduction}\label{sec:intro}

The Relativistic Heavy Ion collider (RHIC) at Brookhaven National Laboratory 
has been taking data from nucleus-nucleus, deuteron-nucleus and proton-proton
collisions starting from 2000. This successful program has provided 
a wealth of information on the properties of QCD matter at high energy density.
 For a review of the early experimental 
results we refer the reader to the experimental ``white 
papers''\cite{Arsene:2004fa,*Adcox:2004mh,*Back:2004je,*Adams:2005dq} 
by the RHIC collaborations.
Since the earliest RHIC observations it has become clear that the produced
matter is deconfined and cannot be understood in terms of phenomenological
models of low energy hadron physics. It is obvious that this will 
be even more true for heavy ion collisions at the LHC.
On the other hand, there are many signs
that the matter interacts far too strongly to be described by naive perturbative 
calculations.
The first principles method of studying 
nonperturbative QCD is to directly discretize the path integral on 
a lattice and evaluate it numerically. This is indeed the method 
of choice for the description of a static system, such as studying the
equation of state. The lattice, however, is ill suited for understanding 
time dependent phenomena, and other methods must be developed.
One is then faced with a choice between two views,
depending on whether ``strongly interacting'' in the context of 
gauge theory implies a large value of the coupling constant $\as$ or not.
One option is that one should consider $\as$ to be large.
Performing practical computations in this strong coupling limit
has now become possible at least in some nonabelian gauge theories, 
if not yet in QCD, due to a duality transformation that makes 
it possible to translate the problem into solving
a classical wave equation in 5~dimensional 
gravity\cite{Klebanov:1999ku,*Aharony:1999ti,*Klebanov:2000me,*Son:2007vk}.
There has been much activity in this direction recently, and gauge/gravity
duality seems a promising tool to gain qualitative understanding  
of the properties of the medium in the plasma phase.

 A stong coupling theory
fails, however, in describing some basic features of high energy hadronic 
collisions that characterize the initial stage of a heavy ion collision.
The predominance of forward scattering in the angular distribution (i.e. 
the approximate boost invariance of particle production, the 
small baryon stopping in hadronic collisions, the existence of jets) 
point towards an interaction that is well described
by a weak coupling gauge 
theory (see e.g. the discussion 
in\cite{Kovchegov:2009he,*Iancu:2008sp,*Iancu:2008er,*Iancu:2008sn}).
This leads to the point of view that we will assume in this paper; namely
that for the initial conditions of heavy ion collisions the system
can be described assuming that $\as$ is small. In spite of the coupling constant
being small the system can be strongly interacting because it is dense; 
gluonic states have high occupation numbers and the interactions
between them are highly nonlinear.
One possible application of this idea is to assume the existence of 
a geometrical condition that restricts the number of 
particles in the central rapidity region when they begin to overlap 
in phase space. In a  practical phenomenological application
 one can then first calculate the  initial pruduction of partons  
using a framework developed for a generic hadronic collision
and then supplement this calculation with the geometric
(``final state'') saturation criterion. Examples of this line of
reasoning are provided e.g. by the
 EKRT model\cite{Eskola:1999fc,*Eskola:2000xq,*Eskola:2002qz,*Eskola:2001bf},
which uses the usual collinear pQCD factorization, or
the parton percolation 
model\cite{Armesto:1996kt,*Nardi:1998qb,*Satz:2002qj,*Digal:2002bm,*Digal:2003sg,*Braun:1997ch},
based on string-like phenomenology.
The line of thought that we will pursue here, however, is that 
the nonlinear interactions that control the behavior of the 
bulk of particle production are present already in the initial 
nuclear wavefunction\cite{Gribov:1984tu,*Mueller:1985wy,*Blaizot:1987nc}
and lead to the emergence of an (``initial state'') 
\emph{saturation scale}; a transverse momentum scale that
then manifests itself in particle production in the collision.

In collisions of protons and nuclei the typical values of $x$ that are probed in the 
wavefunction are $x \sim \ptt/\sqrt{s}$, where $\ptt$ is s typical transverse
momentum scale of the particles being produced. Let us consider, in the center-of-mass
frame of the collision, the nucleus moving in the $+z$ direction. The parton 
with momentum fraction $x$ will have longitudinal momentum 
$p^+ \sim x \sqrt{s}/A$ and will  thus probe the other, 
leftmoving, nucleus at a length scale  $\Delta x^-  \sim A/ (x \sqrt{s})$.
The  longitudinal size of the leftmover is Lorentz-contracted 
from $\ra \sim A^{1/3} \rp$ to $\sim A^{1/3} \rp ( A m_N/\sqrt{s})$.
We see that if $x \ll 1/ (A^{1/3} \rp  m_N)$, the
partons in the rightmoving nucleus will not be able to resolve the  
individual nucleons 
of the leftmoving one. The whole nucleus must therefore be treated as one coherent
target, not as a collection of independent nucleons.

The observation that the large $x$ localized, valence-like, degrees of
freedom are not resolved in the collision, but only the smaller $x$
partons that they radiate, naturally leads to the idea of treating the
two separately in an effective field theory approach. This effective
field theory is known as the Color Glass Condensate (CGC)
\cite{Jalilian-Marian:1997xn,Jalilian-Marian:1997jx,*Jalilian-Marian:1997gr,*Jalilian-Marian:1997dw,*JalilianMarian:1998cb,*Weigert:2000gi,*Iancu:2000hn,*Iancu:2001md,*Ferreiro:2001qy,*Iancu:2001ad,*Mueller:2001uk}
(see\cite{Iancu:2002xk,*Iancu:2003xm,*Weigert:2005us} for reviews 
and\cite{Gyulassy:2004zy} for a summary of the case for the CGC based on
RHIC data.). 
 The CGC describes a high energy hadron in terms of a
color field (the small $x$ gluons) radiated by an
effective color current (the large $x$ degrees of freedom). At high 
energy (small $x$) the radiation is enhanced by large logarithms of 
the energy and the number of gluons is large. When the occupation numbers 
of gluonic states in the wavefunction becomes large enough, $\sim 1/\as$,
the field can be described as a classical one, radiated by classical
effective color charges. The
classical color charges are stochastic random variables with a
probability distribution $W_x[\rho]$.  Both the properties of the initial 
stage of a 
heavy ion collision and observables in DIS at small
$x$ can be computed in terms of these same classical gluon fields.
The color charge distribution depends on nonperturbative input and
cannot completely be computed from first principles. Its dependence on
the energy scale (rapidity) that separates the large and small $x$
degrees of freedom can, however, be computed and expressed in terms of
a renormalization group equation.  The distribution of color charges
is a universal object; it can be measured in one process (ideally DIS)
and then used as an independent input to make prediction for another
one (say, the initial field configurations in a heavy ion collision).
In this sense the situation is analogous to collinear factorized
perturbation theory; there is a universal, nonperturbative distribution
(color charge distribution or parton distribution function), a
separation scale (rapidity or virtuality) and a renormalization group
equation derived from first principles that describes the dependence
of the nonperturbative input on this separation scale.

\section{Glasma initial state} \label{sec:glasma}

We now turn to the basic features of the  glasma\cite{Lappi:2006fp} 
fields in the initial stages of the collision. We shall show hot their
 structure as initially longitudinal boost invariant fields
follows directly from the CGC description of the
nuclear wavefunction.

\subsection{Description of a nucleus in the CGC}\label{subsec:1nucl}

In the CGC picture the small $x$ degrees of freedom that 
contribute to bulk gluon production are treated
as a classical color field radiated by (classical) color
sources. As classical fields they obey the Yang-Mills equation 
of motion
\begin{equation}\label{eq:eom}
[D_{\mu},F^{\mu \nu}] = J^{\nu},
\end{equation}
with the source currents 
\begin{equation}\label{eq:twonucl}
 J^{\mu} =  \delta^{\mu +}\rho_{1}(\xt,x^-) 
+ \delta^{\mu -}\rho_{2}(\xt,x^+)
\end{equation}
representing the large $x$ part of the wavefunctions.
The color charge densities are \emph{static} in the sense
that they do not depend on the relevant light cone time
($x^\pm$ for  $\rho_{1,2}$ respectively). The color charges
of the two nuclei are naturally independent of each other. 
They are also localized on their respective light cones:
$\rho_{1,2}(\xt,x^\mp) \sim \rho_{1,2}(\xt)\delta(x^\mp)$. 
A naive way to understand this dependence is to argue that
at high energy the nuclei are Lorentz-contracted to
infinitesimal sheets in an interval\footnote{We're
concentrating for the moment on the rightmoving nucleus.} 
$\Delta x^- \sim 1/\sqrt{s}$.
This argument does not, however, take into account the quantum 
nature of the color charged integrated into the effective
description in terms of the $\rho$'s. The color charge 
densities represent degrees of freedom that have a higher
$x$, i.e. a higher longitudinal momentum $p^+$
than the ones described as the classical field. Consequently
they are better localized in $x^-$ (the conjugate variable to 
$p^+$) than the classical field and are, in 
the high energy (multi-Regge) kinematics we are working in, 
seen by the glasma field as infinitesimal sheets.
The actual scale of this delta function can be estimated as 
follows. The longitudinal (and transverse) momenta of the 
gluons in the classical field are\footnote{We are assuming that
one is here interested in the field around $y=0$, otherwise
all the scales in the following argument must be boosted
appropriately.} $\sim \qs$. The evolution speed in 
rapidity is proportional to $\as$; we can therefore assume
that the sources are separated from the classical field
by $\Delta y \sim 1/\as $ units in rapidity. Thus
they correspond to longitudinal momenta $p^+ \sim e^{1/\as} \qs$, 
i.e. are localized in an interval $\Delta x^- \sim e^{-1/\as} /\qs$.
This is the actual meaning of the delta function approximation 
used when writing $\rho_{1}(\xt,x^-) \sim \rho_{1}(\xt)\delta(x^-)$.
As we shall see in the following, it is nevertheless important 
to initially maintain an explicit $x^-$-dependence in the color charge 
density and only take the $\delta$-function limit later at the 
appropriate stage. In the leading logarithmic kinematics 
the $x^-$-dependence will be 
identified with the dependence of the color charge density on
the rapidity (or $x$) at which it is probed.

The computation of the glasma fields from the current 
\nr{eq:twonucl} proceeds in the following way\cite{Kovner:1995ts}.  
One must first solve the problem for the field of one individual nucleus;
the gauge fields of the individual nuclei will then give the initial condition
for computing the field configuration in a two-nucleus collision.
We shall first go through these small manipulations that will give 
us a picture of the initial glasma field configurations after the
collision before discussing the relation to the renormalization 
group evolution of the sources.

The solution for the color current of one nucleus
\begin{equation}\label{eq:onenucl}
 J^{\mu} =  \delta^{\mu +}\rho_{1}(\xt,x^-) 
\end{equation}
is most 
easily found in the covariant gauge $\partial_\mu A_\mathrm{cov}^\mu=0.$
One can find a solution with only one component of the gauge field is 
nonzero, namely $A_\mathrm{cov}^+(\xt,x^-).$ In this case \eq\nr{eq:eom} becomes 
a 2-dimensional Poisson equation
\begin{equation}
-\nabt^2 A_\mathrm{cov}^+ = \rho(\xt,x^-),
\end{equation}
for which we can formally write the solution as
\begin{equation}\label{eq:apluscov}
A_\mathrm{cov}^+ = - \rho(\xt,x^-) /  \nabt^2.
\end{equation}
Note that there is an infrared singularity in \eq\nr{eq:apluscov}. The most
natural prescription to invert the Laplacian $\nabt^2$
is to impose the constraint 
$\int \ud^2 \xt \rho(\xt)=0,$ i.e. to require that the source as a whole
is color neutral. Imposing color neutrality at a shorter length scale will
also remove this ambiguity.

The covariant gauge solution has the advantage of being localized on the light 
cone in the $t,z$-plane, but its interpretation in terms of partons is not
very clear. To interpret the classical field in terms of 
quasi-real Weizs\"acker-Williams gluons we must transform the field into 
the light cone (LC) gauge. This gauge transformation can be done
 using the path ordered exponential
\begin{equation} \label{eq:pathorder}
U(\xt,x^-) = \mathrm{P} \exp \left\{ ig \int_{-\infty}^{x^-} \ud y^-
A_\mathrm{cov}^+(\xt,y^-)
\right\},
\end{equation}
giving
\begin{eqnarray} \label{eq:LCsol}
A^\pm_\mathrm{LC} & = &  0 \\
A^i_\mathrm{LC} & = & \frac{i}{g} U(\xt,x^-) \partial_i U^\dag(\xt,x^-).
\end{eqnarray}
The light cone gauge solution is not localized on the $x^+$-axis, unlike the
one in covariant gauge. Instead, it extends to the whole region
$x^- >0$ as a transverse pure gauge field.
The field strength tensor $F^{\mu \nu},$ however is nonzero only on the light
cone $x^-=0$ because there is a nonzero energy density only following the
color source.

In general the color charge densities are stochastic
random variables drawn from some distribution $W_y[\rho]$. This 
probability distribution cannot be computed from first principles,
but its dependence on $y$ can; it is given by the JIMWLK
renormalization group equation, which 
we shall return to shortly. 
For many phenomenological applications and the following
discussion, it will be enough to consider the distribution
defined by the McLerran-Venugopalan (MV) model\cite{McLerran:1994ni},
which could be considered as a reasonable
initial condition for JIMWLK.
In the MV model
the color charges are distributed with the Gaussian 
probability distribution
\begin{equation}\label{eq:korr}
\langle \rho^a(\xt,x¯) \rho^b(\yt,y¯) \rangle = g^2 \mu_A^2 
\delta^{ab}\delta^2(\xt-\yt) \delta(x^- - y^-).
\end{equation}
One can motivate the simple Gaussian approximation by an argument 
that is  essentially based on the central limit theorem. The 
classical color charge density in the CGC is conceptually 
a sum of the color charges of all the higher $x$
partons that have been integrated out of the theory. If these
color charges are uncorrelated (incoherent) and there are sufficiently many 
of them, the resulting distribution will be 
Gaussian\cite{Jeon:2004rk} and local in transverse space.
This assumption of independent color charges adding up to the 
effective charge $\rho(\xt)$ also leads to the assumption
that the parameter $\mu^2$ should be proportional to the thickness 
of the target (i.e. $\mu^2 \sim A^{1/3}$); a dependence that
is modified by quantum evolution that introduces correlations
between the charges\cite{Mueller:2003bz,Kowalski:2007rw}.
The MV model has two important properties that we must mention here.
Firstly, as originally shown in Ref.\cite{Jalilian-Marian:1997xn},
 it leads to \emph{saturation} of the unintegrated gluon 
distribution (and of the dipole cross section), at transverse
momentum scales $\ktt \lesssim \qs \sim g^2\mu$
(we shall comment on the precise relation between
$g^2 \mu $ and $\qs$ later, see also 
\cite{Fukushima:2007ki,Lappi:2007ku}). Secondly, in the 
dilute large transverse momentum limit it leads to an unintegrated
gluon distribution that behaves as $\sim 1/\ktt^2$. In terms of the 
integrated distribution this means that $xG(x,Q^2) \sim \ln Q^2$, 
the kind of generic  behavior one expects from DGLAP evolution.
These two features; saturation and a perturbative-like gluon spectrum
at large $\kt$, make it reasonable to expect that the MV model 
should be a realistic one for phenomenology at RHIC energies, 
where the effects of JIMWLK evolution have not yet drastically modified
the distribution of the color charges.

\subsection{Glasma fields}\label{subsec:2nucl}

Let us then turn to the case of two colliding nuclei. It is more convenient 
here to work in an axial gauge. The reason for this is that we would like 
to understand the solution of the classical equations of motion inside the
future light cone $\tau>0$ as an \emph{initial value problem} involving
only the initial conditions of the fields $A_\mu$ at $\tau=0$ and subsequent 
propagation into the vacuum. If one studies the solution of the
two nucleus problem in the covariant gauge the large $A^+$-component 
of the \emph{field} of the rightmoving source causes a precession (due to 
covariant current conservation $[D_\mu,J^\mu] = 0$) of the \emph{current}
of the leftmoving nucleus. Thus in the covariant gauge an essential part
of the interaction involves the currents directly. In a light cone gauge
one can take advantage  of the fact that the solution
for the one nucleus case has, besides the gauge condition
 $A^+=0$, also the 
property\footnote{This is for the rightmoving nucleus, for the leftmover
$A^-=0$ is the light cone gauge condition and $A^+=0$ a property of the
particular solution.} $A^-=0$. This is due to the static nature of the current;
a $x^+$-dependence in the Wilson line \nr{eq:pathorder} would induce a 
$A^-$-field. In other words this is a feature of the leading $\ln x$ kinematical 
regime, where the degrees of freedom in the source are assumed to have 
a large $p^+$, virtuality $\sim \ptt^2 \sim \qs^2$ and therefore a very small $p^-$.
Choosing purely $A^+=0$ or $A^-=0$ as our gauge condition would, however,
break the symmetry between the two nuclei in an inconvenient way
(see, however\cite{Blaizot:2008yb} for a discussion of the CYM problem
in an asymmetric gauge and\cite{Altinoluk:2009je,Altinoluk:2009jf}
for a formulation of the problem in a way that is completely asymmetric
from the beginning). Fortunately a convenient
symmetric axial gauge condition is provided by the temporal, or Fock-Schwinger,
gauge in the proper time-rapidity coordinates: $A_\tau=0$. The other
independent linear combination, $A_\eta = -\tau^2 A^\eta = x^+ A^- - x^- A^+$
is zero for the one-nucleus solution, but not in the case of two 
colliding nuclei. The Schwinger 
gauge condition in light cone coordinates is
$x^+A^- + x^+A^-=0$. In other words, on the $x^+=0$-light cone where there
is a current $J^-$, the gauge condition guarantees that $A^+=0$ and the
current does not precess, and symmetrically for the other light cone.
The effects of the interaction between the two nuclei are now visible only 
in the gauge field, not the current (at least at the classical level).

\begin{figure}[tbp]
\begin{center}
\includegraphics[width=0.70\textwidth]{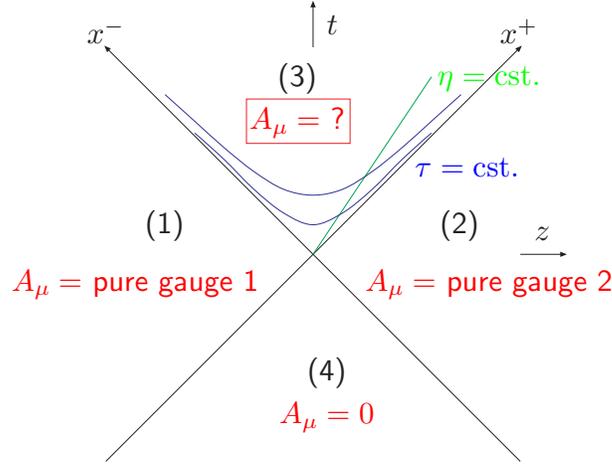}
\end{center}
\caption{Color fields in spacetime. In regions (1) and (2), where only one
of the nuclei has passed by, the field is the pure gauge field of this one
nucleus. In region (3) the field is known numerically.}
\label{fig:spacet}
\end{figure}

Inside the future light cone (the spacetime structure is illustrated
 in \fig\ref{fig:spacet}) the gauge 
fields satisfy the equations of motion in vacuum. What is needed is the initial
condition for solving these equations. These initial conditions can be obtained
by requiring that the fields in the different regions match smoothly on the 
light cone. In practice this is done  by inserting
the ansatz
\begin{eqnarray} \label{eq:iniansatz}
A^i &=& \theta(-x^+)\theta(x^-) A^i_{(1)}
+ \theta(x^+)\theta(-x^-) A^i_{(2)}
+\theta(x^+)\theta(x^-) A^i_{(3)} \\
A^\eta &=&  \theta(x^+)\theta(x^-) A^\eta_{(3)}
\end{eqnarray}
into the equation of motion \nr{eq:eom} and requiring that the singular terms
arising from the derivatives of the $\theta$-functions cancel. 
In this way one gets the following initial conditions
for the gauge field in the future light cone:
\begin{eqnarray}\label{eq:trinitcond}
A^i_{(3)}|_{\tau=0} &=& A^i_{(1)} + A^i_{(2)} \\
\label{eq:longinitcond}
A^\eta_{(3)}|_{\tau=0} &=& \frac{ig}{2}[A^i_{(1)},A^i_{(2)}].
\end{eqnarray}
The equations of motion with these initial conditions can then be solved 
either numerically or perturbatively in the weak field limit. 
We shall turn to these solutions later, but the basic features of
the glasma fields can be seen already from the initial conditions.

\begin{figure}[tbp]
\begin{center}
\includegraphics[width=0.8\textwidth]{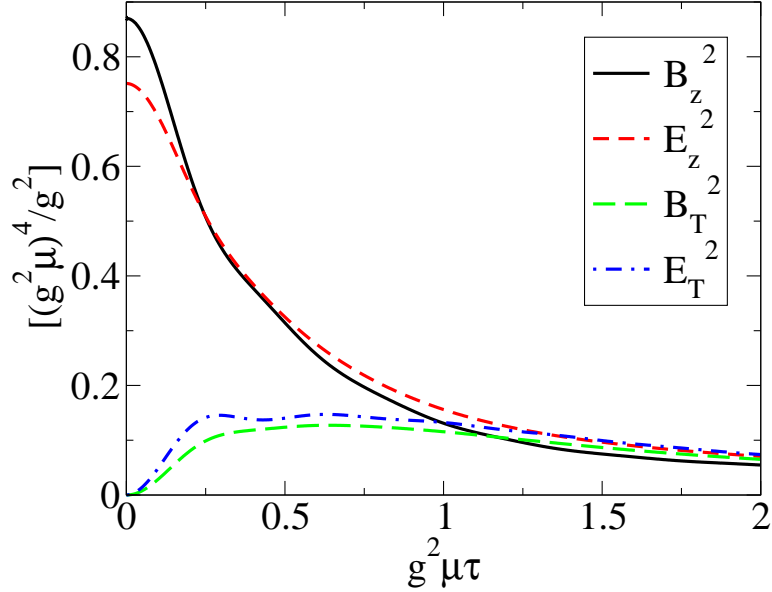}
\end{center}
\caption{Components of the gauge field, computed numerically on a 
$512^2$-lattice with $g^2\mu\ra = 67.7$.}
\label{fig:fields}
\end{figure}

For a physical interpretation it is useful to calculate the corresponding
chromoelectric and -magnetic fields in the usual $t,z$-coordinate system.
The field configurations \eqs\nr{eq:trinitcond} and~\nr{eq:longinitcond}
correspond to vanishing transverse fields and longitudinal components 
given by
\begin{eqnarray}
\label{eq:longite}
	E^z & = & ig[A^i_{(1)},A^i_{(2)}] \\
\label{eq:longitb}
    B^z & = & ig \epsilon^{ij} [A^i_{(1)},A^j_{(2)}].
\end{eqnarray}
Note that the gauge potential $A_\mu$ is a true vector and therefore the 
longitudinal component $A_\eta$ corresponds to $A_z$ only at $\eta=0$. The
field strengths $\vec{E}$ and $\vec{B}$, on the other hand are components
of the tensor $F^{\mu\nu}$ and \eqs\nr{eq:longite} and~\nr{eq:longitb} are true
for all $\eta$. Solving the equations of motion forward in time then 
generates also transverse components for the fields within a time $\sim 1/\qs$.
A plot of the transverse and longitudinal color field strengths 
as a function of $\tau$ from a numerical calculation  is shown in 
\fig\ref{fig:fields}.

\begin{figure}
\includegraphics[width=0.45\textwidth]{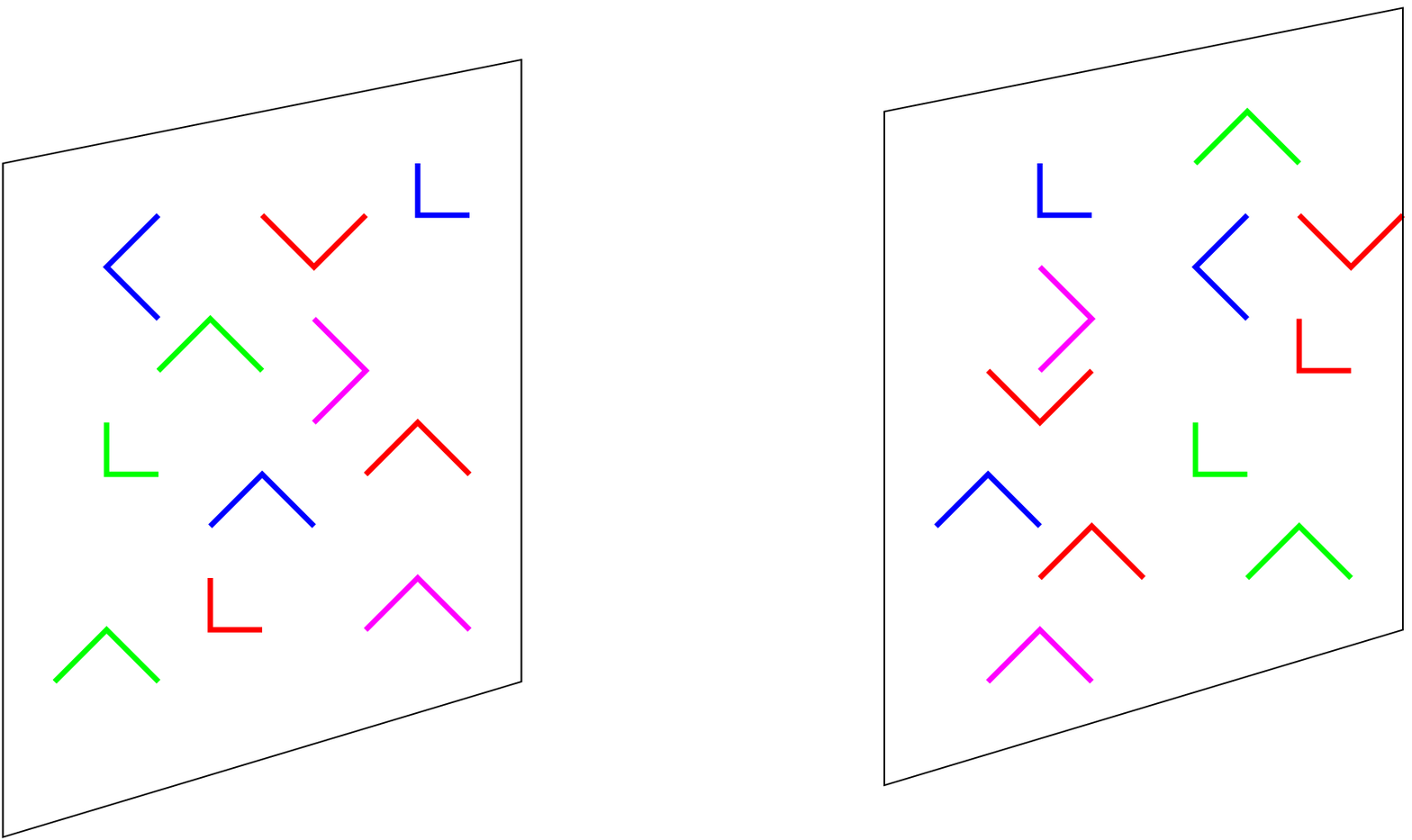}
\hfill
\includegraphics[width=0.45\textwidth]{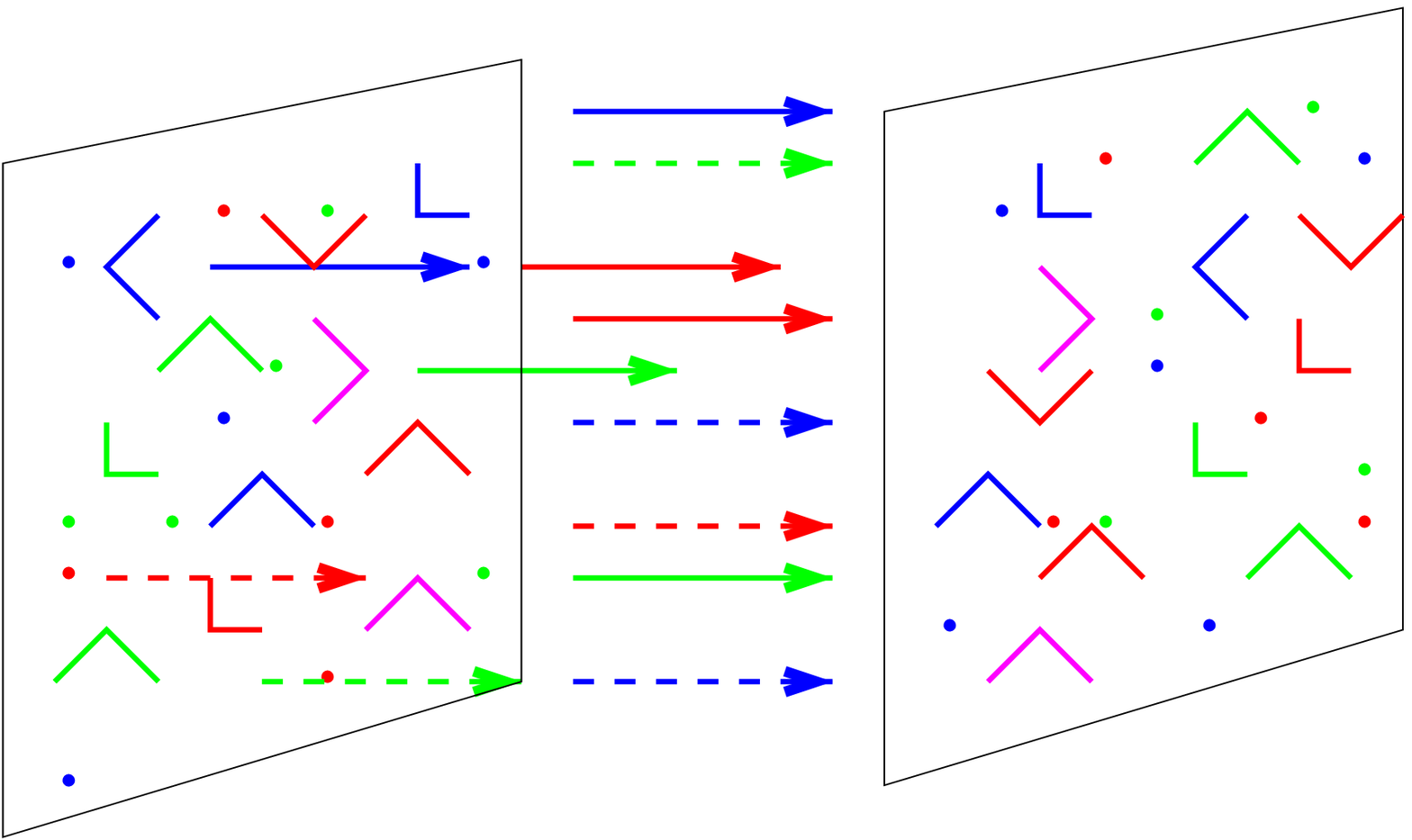}
\caption{The WW fields of the two nuclei before and after
the collision. Before the collision there are only transverse fields
on the sheets. After the collision the interaction of these fields
with the pure gauge field of the other nucleus leaves behind 
an effective electric and magnetic charge density (the dots on the figure) 
on the sheet, and a longitudinal electric and magnetic field between these
effective  charges.}
\label{fig:sheetonsheet}
\end{figure}

The color fields of the two individual nuclei are transverse electric and magnetic 
fields on the light cone. 
Why then are the glasma fields in the region between the two nuclei 
longitudinal along the beam axis? One way of 
understanding these field configurations is the following.
Let us work still in the $A_\tau=0$ gauge, so that each nucleus, when going past
a point on the beam axis with no gauge field before the collision, leaves
behind it a pure gauge field (see \fig\ref{fig:spacet}).
One can define an effective chromoelectric and chromomagnetic charge density by 
separating the nonlinear parts of the vacuum 
Gauss law and Bianchi identities
\begin{equation}
\left[ D_i,E^i \right] = 0 \quad \textrm{ and } \quad 
\left[ D_i,B^i \right] = 0 
\end{equation}
as
\begin{equation}
\partial_i E^i  = \rho_\mathrm{e} =  i g [A^i,E^i] \quad \textrm{ and } \quad
\partial_i B^i = \rho_\mathrm{m} = i g [A^i,B^i].
\end{equation}
Now we can interpret the interaction of the WW chromoelectric and -magnetic 
fields of the nucleus on the $x^+$-light cone with the pure gauge field left
 behind by the other nucleus as an effective chromoelectric
and -magnetic charge density left behind on the light cone. An exactly
opposite charge density is left behind on the other sheet, leading to a 
longitudinal chromoelectric and -magnetic field between the sheets. This
structure is illustrated in \fig\ref{fig:sheetonsheet}. 
This description is
very similar in spirit to the original suggestions for a string based
description of a hadronic collisions. These string picture had only sources
of electric charge~\cite{Andersson:1983ia,%
*Ehtamo:1983hu,*Biro:1984cf,*Gatoff:1987uf}. When these longitudinal field 
configurations are actually derived from the QCD equations of motion
also a magnetic field naturally appears.

It is interesting to note the structure of the energy momentum tensor 
$T_{\mu \nu} = \frac{1}{4}g_{\mu \nu}F^{\alpha \beta}F_{\alpha\beta}
- F_\mu{}^\alpha F_{\nu \alpha} $ for this initial condition. It is diagonal
and, as always in gauge theory at the classical level, traceless: 
$T_{\mu \nu} = \half (E_z^2 + B_z^2) \times \mathrm{diag}(1,-1,-1,1)$. This can be
compared to the standard form for a system with an anisotropy in the $z$-direction:
$T_{\mu \nu}= \mathrm{diag}(\epsilon,-p_\perp,-p_\perp,-p_L),$
where $\epsilon$ is the energy density and
$p_\perp$ and $p_L$ are the transverse and longitudinal pressures. We see
that the initial field configuration has \emph{negative} ``longitudinal pressure''.
The configuration that is the starting point for studies of isotropization
by plasma instabilities, where the diagonal elements of $T_{\mu\nu}$ at
$\eta=0$ are $(\epsilon,-\epsilon/2,-\epsilon/2,0)$, is only reached at times
$\tau \gtrsim 1/\qs$ when the classical fields start to behave linearly due
to the expansion of the system.

The glasma field configurations depend on the transverse coordinate at
the length scale $1/\qs$, which corresponds to the typical transverse 
momentum of the gluons being $\sim \qs$. The same scale is also in 
general the correlation length of the system, and will as such determine
the strength of multigluon correlations as we shall see in 
\se\ref{sec:corr}. This leads to the lifetime of the 
purely longitudinal field configuration also being $\sim 1/\qs$. For the 
bulk of particle production there is therefore no clear separation of timescales 
that would justify treating the field as constant in time and space,
no matter how appealing this approximation migh be for those used to 
low energy string phenomenology. The difference between the glasma and 
the Lund string model is that conceptually the transverse scale
of the problem is a semi-hard scale $\qs$, not the confinement scale 
$\lqcd$. Note also that the 
initial fields being longitudinal along the beam axis direction
is in no contradiction with the lowest order perturbative description
of the process as $gg \to g$ scattering, because the longitudinal 
(with respect to the beam axis) fields are perpendicular
to the momentum  of the gluon being produced. The initial
polarization state of this gluon is, however, a very particular one.

The field
inside the future light cone can then be computed either
numerically~\cite{Krasnitz:2001qu,Lappi:2003bi,Krasnitz:2003jw} or
analytically in different approximations (see
e.g.~\cite{Blaizot:2008yb} for recent work).  The obtained result is
then averaged over the configurations of the sources $J^\mu$ with a
distribution $W_y[J^\mu]$ that includes the nonperturbative knowledge
of the large $x$ degrees of freedom.  The resulting fields are then
decomposed into Fourier modes to get the gluon spectrum.
  This is the method that we will refer to as
Classical Yang-Mills (CYM) calculations.  Note that the average over
configurations is a classical average over a probabilistic
distribution. This is guaranteed by a
theorem~\cite{Gelis:2008rw,Gelis:2008ad} ensuring the factorization of
leading logarithmic corrections to gluon production into the quantum
evolution of $W_y[J^\mu]$, analogously to the way leading logarithms
of $Q^2$ are factorized into DGLAP-evolved parton distribution
functions.

In the limit when either one or both of the color sources are dilute
(the ``pp'' and ``pA'' cases), the CYM calculation can be done
analytically and reduces to a factorized form in terms of a
convolution of unintegrated parton distributions that can include
saturation effects:
\begin{equation}
\frac{\ud N}{\ud^2\pt \ud y} =
\frac{1}{\as}\frac{1}{\pt^{\!\!\!2}} \int
 \frac{\ud^2\kt}{(2 \pi)^2} 
\varphi_y(\kt) \varphi_y(\pt - \kt).
\end{equation}
  Although this approach (known as ``KLN'' after
the authors of~\cite{Kharzeev:2000ph,Kharzeev:2001gp}, 
see\cite{Gribov:1984tu} for the original work) is not strictly
valid for the collision of two dense systems, it does have the
advantage of offering some analytical insight and making it easier to
incorporate large-$x$ ingredients into the calculation.
We will discuss these calculations of the gluon spectrum in more
 detail in \se\ref{sec:bulk}.

\subsection{Factorization}
\label{subsec:fact}

In the previous discussion we did not specify the probability distribution
of the color charge densities more precisely, besides mentioning
the MV model as a phenomenological approximation. The color charge 
distribution $W_y[\rho]$ includes nonperturbative information 
about the large-$x$ part of the nuclear wavefunction, i.e. the valence 
degrees of freedom boosted from the rest frame of the nucleus.
Thus the probability distribution cannot be computed from first
principles in weak coupling. Nevertheless, by considering quantum corrections
to high energy scattering, it is possible 
to derive a \emph{renormalization group} equation that is known
by the acronym JIMWLK\cite{Jalilian-Marian:1997xn,Jalilian-Marian:1997jx,*Jalilian-Marian:1997gr,*Jalilian-Marian:1997dw,*JalilianMarian:1998cb,*Weigert:2000gi,*Iancu:2000hn,*Iancu:2001md,*Ferreiro:2001qy,*Iancu:2001ad,*Mueller:2001uk}
(pronounced as ``gym-walk''). The JIMWLK equation describes
the dependence of the probability distribution $W_y[\rho]$ on 
$y$, the rapidity (or, equivalently in the leading log high energy
kinematics that we are working in, on $\ln 1/x$). In a mean-field
approximation valid in the large $\nc$-limit the JIMWLK evolution
of the correlation function of two Wilson lines reduces to 
the BK\cite{Balitsky:1995ub,Kovchegov:1999ua,*Kovchegov:1999yj}
equation.

The role of  $W_y[\rho]$ is analoguous to the conventional 
parton distribution function; it is a nonperturbative quantity
whose dependence on one of the kinematical variables of the 
process is described by a weak coupling renormalization group equation.
In the case of parton distributions, the renormalization group equations
are the DGLAP ones, and they describe evolution in $Q^2$.
In the case of DGLAP one is dealing with a dilute system, where
the appropriate degrees of freedom are individual partons
with a definite momentum, whereas in the case of the CGC 
the good degrees of freedom are color charges resulting
from interactions of many partons. The distributions
$W_y[\rho]$ are similar to parton distributions in the
sense that they are not (complex) wavefunctions but (at 
least loosely speaking) real probability distributions. 
This is guaranteed by \emph{factorization theorems}
stating that there is no interference between the 
dynamics of the hadrons or nuclei 
at large $x$ (JIMWLK; or BFKL) or at small $Q^2$ (DGLAP)
and the process one is studying.
Factorization can be understood as a statement that 
one has found the right set of degrees of freedom 
in which one can compute physical observables from only the diagonal
elements of the density matrix of the incoming nuclei.

In our nonlinear high-energy context the JIMWLK equation
was derived in the context of deep inelastic scattering
off a nuclear target. In the case of particle production in the glasma,
it was shown more recently~\cite{Gelis:2007kn,Gelis:2008rw} that 
when one computes the NLO quantum corrections
to a given observable in the Glasma, all the leading 
logarithmic divergences can be absorbed into the RG evolution of the sources
with the same Hamiltonian that was derived by considering only the DIS process. 
The underlying physical reason for factorization is that this
fluctuation with a large $k^+$ requires such a long interval in $x^+$ 
to radiated that it must be produced well before and independently of the
interaction with the other 
(left moving and thus localized in $x^+$) source.

In the case of the spectrum of gluons produced in a high energy collision
the leading log divergence can be written as
two JIMWLK Hamiltonians, one for each nucleus,
 acting on the expression for the leading order spectrum.
This same Hamiltonian
describes the RG evolution of the source distributions $W_y[\rho]$.
It is most naturally expressed as
\begin{equation}\label{eq:hjimwlk}
\mathcal{H}
\equiv \frac{1}{2}
\int \ud^2\xt \ud^2\yt 
D_a(\xt)
\eta^{ab}(\xt,\yt)
D_b(\yt)
\end{equation}
in terms of Lie derivatives $D_a(\xt)$ operating on
 the Wilson lines, \eq\nr{eq:pathorder},  constructed from the source color charge densities. 
The kernel in \eq\nr{eq:hjimwlk} is a function of these same Wilson lines:
\begin{multline}
\eta^{ab}
(\xt,\yt)
=
\frac{1}{\pi}
\int \ud^2\ut
\frac{(\xt-\ut)\cdot(\yt-\ut)}{(\xt-\ut)^2(\yt-\ut)^2}\;
\Big[
U(\xt) U^\dag(\yt)
\\
-U(\xt) U^\dagger(\ut)
-U(\ut) U^\dag(\yt)
+1\Big]^{ab} 
\label{eq:eta-f}.
\end{multline}
The fact that no other divergent terms appear is the 
proof of factorization; this is the central result of Ref.\cite{Gelis:2008rw}.

\section{Nuclear wavefunction}\label{sec:wavef}

\begin{figure}[t]
\begin{center}
\noindent
\includegraphics[width=0.7\textwidth]{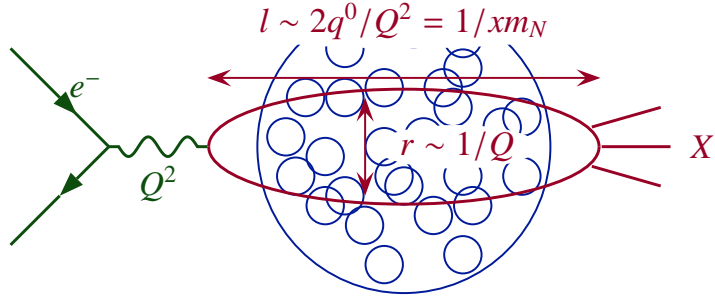}
\end{center}
\caption{In the dipole frame the incoming virtual photon splits into a 
quark-antiquark dipole of trensverse size $\rt$, which then interacts with the
target with the dipole cross section $\hat{\sigma}$.
}\label{fig:dipole}
\end{figure}

\subsection{Deep inelastic scattering}\label{subsec:eA}
The most direct probe of the nuclear wavefunctions would be to perform 
nuclear deep inelastic scattering experiments at high energy such as at
the  Electron-Ion-Collider (EIC)~\cite{Deshpande:2005wd} or
a Large Hadron-electron Collider (LHeC)~\cite{Dainton:2006wd}. In the meantime
one has to rely on the few existing, relatively low energy, measurements and
theoretical extrapolations from proton data from HERA. For a review of the existing
understanding of nuclear shadowing  we refer
the reader to the review by Armesto~\cite{Armesto:2006ph}.

\begin{figure}[t]
\begin{center}
\resizebox{\textwidth}{!}{
\includegraphics[height=5cm]{uspectsc.eps}
\includegraphics[height=5cm]{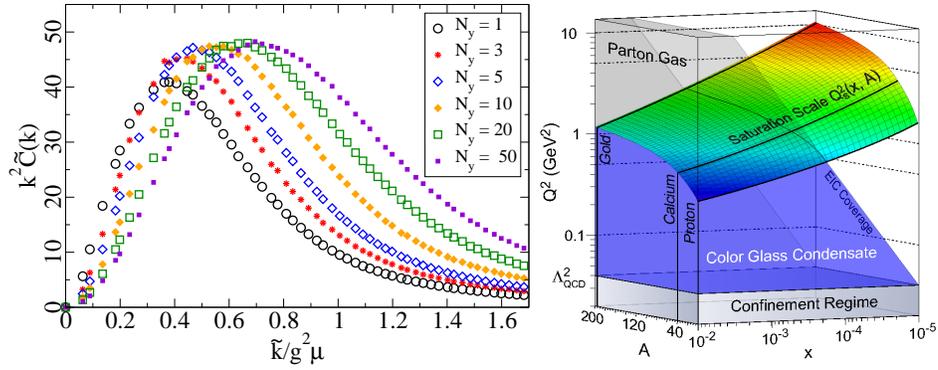}
}
\end{center}
\caption{
Left: Wilson line correlator multiplied  by $\ktt^2$ in the MV model,
with different discretizations of the longitudinal 
coordinate\protect\cite{Lappi:2007ku} (see \eq\nr{eq:uprod} below). 
Reading off the maximum of these curves and defining the corresponding
$\ktt$  as $\qs$  is one possible way to relate the parameters
of the MV model to $\qs$.
Right: estimate of the numerical value of the saturation scale, 
based on a fit to HERA data and a Woods-Saxon parametrization
of the nuclear geometry\protect\cite{Kowalski:2003hm,Kowalski:2007rw}.
}\label{fig:Qs}
\end{figure}

It is useful to think of deep inelastic scattering at small $x$
in the dipole picture \cite{Mueller:1993rr,*Mueller:1994gb,*Mueller:1994jq,
*Andersson:1990dp,*Nikolaev:1990ja,McDermott:1999fa},
where the process is viewed as a virtual 
quark fluctuating into a color dipole, which then probes the wavefunction
of the target (see \fig\ref{fig:dipole}).

In the dipole model one factorizes the total cross section into the
probability for the virtual photon to fluctuate into a $q\bar{q}$ pair
(a color dipole) and the cross section of the dipole scattering
with the target\footnote{There are some tricky issues related to the 
Lorentz frame in which one should view the scattering process in 
the dipole model, see e.g. the discussion in
\protect\cite{Brodsky:2002ue,Brodsky:2004hi}.}.
The total cross section can be written as 
\cite{Golec-Biernat:1998js,McDermott:1999fa}:
\begin{equation}\label{eq:f2}
\sigma_{\mathrm{T,L}}(x,Q^2) = \int \ud^2 \rt \int_0^1 \ud z
|\psi_{\mathrm{T,L}}(z,\rt)|^2 \sigma_\textrm{dip}(x,Q^2,\rt). 
\end{equation}
Here the \emph{photon wave function} $\psi_{T,L}(z,\rt)$ gives the probability
for the virtual photon (T and L stand for, respectively, 
 transverse and longitudinal polarizations
of the photon) to split into a color dipole of transverse size $\rt$. The
wave function $\psi_{T,L}(z,\rt)$ includes the known QED part of the
reaction and is known analytically\footnote{
To leading order in $\alpha_{\mathrm{em}}$, which is quite sufficient
in this context.}. The exact expressions can be found in e.g.~\cite{Golec-Biernat:1998js}.
In the classical field approximation the dipole 
cross section can be expressed in terms
of the Wilson lines $U(\xt)$  appearing in \eq\nr{eq:pathorder}
as\cite{Balitsky:1995ub,Buchmuller:1996xw,*Buchmuller:1998jv,Mueller:2001fv}
\begin{multline} \label{eq:wlinecorr}
\sigma_\textrm{dip}(\rt) = \frac{2}{\nc} \int \ud^2 \bt \Tr  
\left\langle 1 - U^\dag(\bt + \half \rt) U(\bt-\half \rt) \right\rangle
\\
=  2 \int \ud^2 \bt \mathcal{N}(\bt,\rt),
\end{multline}
where we have denoted the imaginary (and dominant at high energy) part of the 
dipole-target scattering amplitude with $\mathcal{N}(\bt,\rt).$
Equation \nr{eq:wlinecorr} provides a direct and explicit connection with 
between DIS observables and the classical field description of the initial
color fields in a nucleus-nucleus collision. For small dipoles the
scattering amplitude is proportional to $\rt^2$, whereas for a large
one it approaches the unitarity limit $\mathcal{N}=1$.
 The saturation scale 
$\qs$ is defined as the characteristic momentum scale separating these two 
regimes. Fourier-transforming the Wilson line correlator w.r.t. $\rt$ into
momentum space $\qs$ could be defined as the $\ktt$ corresponding to the 
maximum of the function $\ktt^2 \mathcal{N}(\bt,\kt)$. In coordinate
space one can define it from the value of  $\rt$ when the scattering 
amplitude reaches some characteristic value. For example one can choose
to  define  $\qs(\bt)$ by the relation\cite{Kowalski:2006hc} 
$\mathcal{N}(\bt,\rtt = \sqrt{2}/\qs) = 1-e^{-1/2}$ or
by\cite{Kowalski:2007rw}  
$\mathcal{N}(\bt,\rtt = 1/\qs) = 1-e^{-1/4}$.
Several other constants in lieu of $1-e^{-1/2}$ or $1-e^{-1/4}$ are used by 
different authors, leading to slightly different numerical values.
The three cited above have the advantage agreeing with the convention
used in the well-known ``GBW'' fit\cite{Golec-Biernat:1998js,Golec-Biernat:1999qd} 
 to HERA data which is discussed below (see \eq\nr{eq:gbw}).
The momentum space Wilson line correlator in the MV model is shown in 
\fig\ref{fig:Qs} (left).

\subsection{Proton-nucleus collisions}\label{subsec:pA}

\begin{figure}[!h]
\begin{center}
\noindent
\includegraphics[angle=270,width=0.7\textwidth]{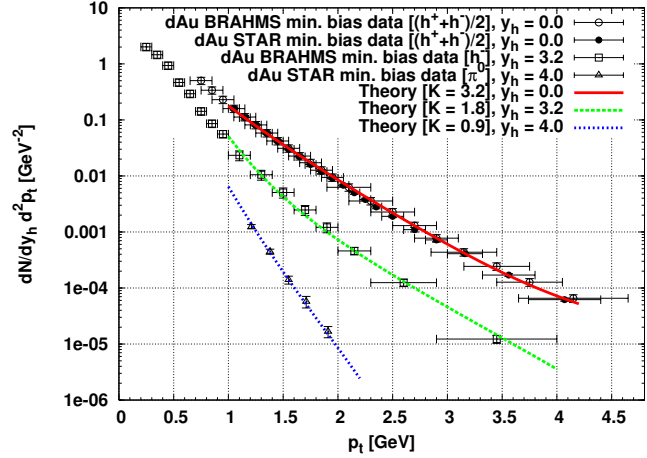}
\end{center}
\caption{
Hard particle spectra in dAu collisions using the DHJ parametrization
of the dipole cross sections\protect\cite{Dumitru:2005kb}.
}\label{fig:pA}
\end{figure}

Another avenue to access properties of the nuclear wavefunction is
to study proton-nucleus collisions and  compare them 
with nucleus-nucleus ones. Here the idea is to treat the proton 
as dilute probe scattering off the CGC of the 
nucleus. Among CGC theorists it is a common terminology 
to denote generally the (formal) dilute-dense limit of scattering by 
``pA''\cite{Gelis:2001da,Dumitru:2001ux,Dumitru:2001jn,Dumitru:2002qt}; many 
of the results are equally valid for forward scattering in AA 
collisions\cite{Gelis:2006tb}. 
Deuteron-nucleus (falling into the same dense-dilute cathegory)  
collisions at RHIC have turned out to be a very powerful
tool to access properties of the small $x$ nuclear wavefunction.
Indeed the 
suppression of high $\ptt$ hadron production in pA collisions at forward
rapidities observed by the BRAHMS collaboration\cite{Arsene:2004ux} 
have been considered
as one of the clearest direct experimental signals from RHIC
favoring gluon saturation. 
Much of the theoretical context and the earlier phenomenologial applications
have been extensively presented in the review 
by Jalilian-Marian and Kovchegov\cite{JalilianMarian:2005jf},
and we shall here discuss them only briefly. 
More recently the focus in pA collisions as
well as nucleus-nucleus ones has been on azimuthal and rapidity correlations;
we shall return to these towards the end of this section.

\begin{figure}
\resizebox{\textwidth}{!}{
\includegraphics[height=5cm]{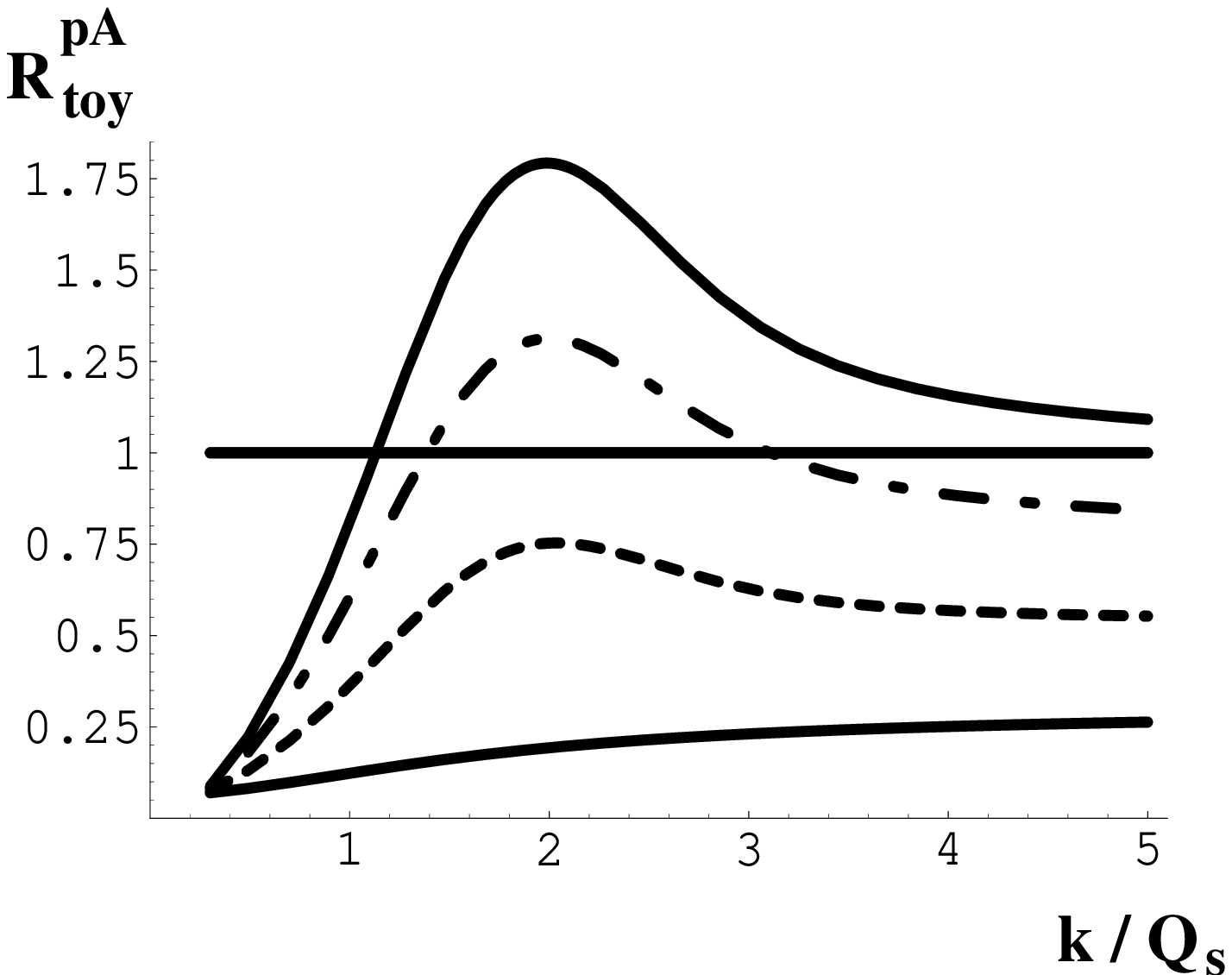}
\includegraphics[height=5cm]{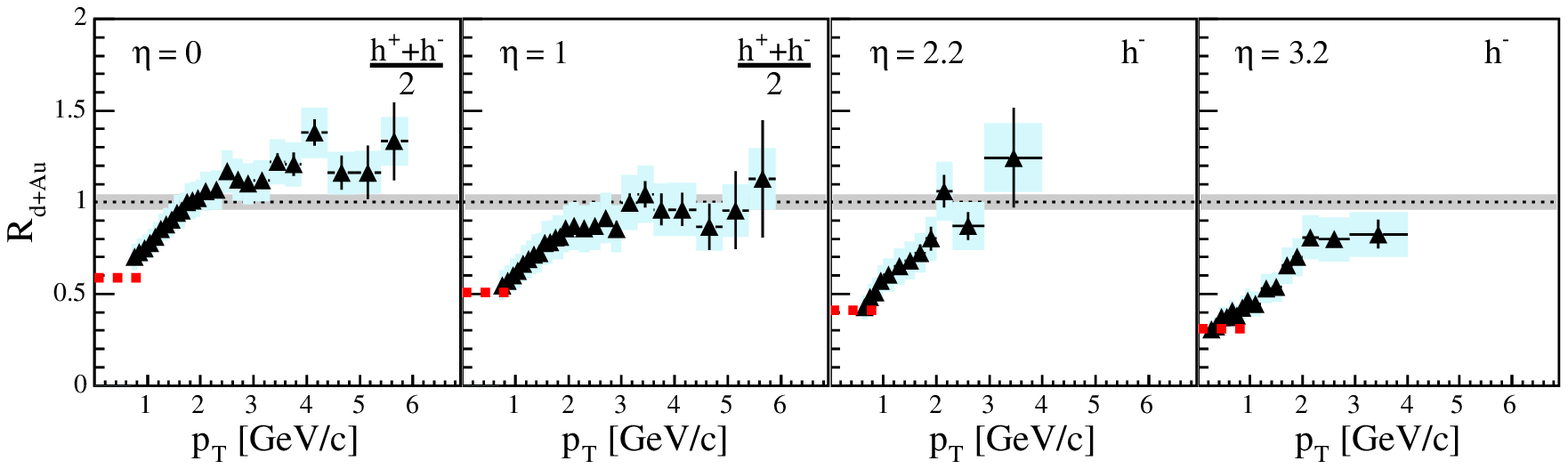}
}
\caption{
Left: nuclear modification factor $\rpa$ from 
BRAHMS\protect\cite{Arsene:2004ux}. 
Right: qualitative prediction of the evolution of $\rpa$
from a CGC calculation\protect\cite{Kharzeev:2003wz}.
}\label{fig:rpa}
\end{figure}

In pA-collisions the particularly interesting kinematical regime is
in the forward rapidity region (proton 
fragmentation region, i.e. small $x$ in the nucleus and large $x$ in 
the proton), where the nuclear saturation scale is large and 
the proton (deuteron in practice) is a dilute probe.
Here one is typically dealing with particle production at
large transverse momenta compared to the intrinsic $\kt$
in the proton wavefunction, and it is necessary to resum the large DGLAP
logarithms on the proton side. One must also take into account both the
quark and gluon degrees of freedom from the proton, whose
scattering off the CGC target depends on the Wilson line correlator
in different representations. The single inclusive hadron spectrum can be
written as\cite{Dumitru:2005gt} 
\begin{multline}\label{eq:pA}
\frac{\ud \sigma^{pA\to h X}}{\ud y \ud^2\pt\ud^2\bt}
=
\frac{1}{(2\pi)^2} \int_{x_\textrm{F}}^1 \ud x_p \frac{x_p}{f_\textrm{f}}
\Big\{
f_{q/p}(x_p,Q^2) \mathcal{N}_\textrm{F}\left[\frac{x_p}{x_\textrm{F}} \pt,\bt \right]
D_{h/q}\left(\frac{x_\textrm{F}}{x_p}\right)
+
\\
f_{g/p}(x_p,Q^2) \mathcal{N}_\textrm{A}\left[\frac{x_p}{x_\textrm{F}}\pt,\bt \right]
D_{h/g}\left(\frac{x_\textrm{F}}{x_p}\right)
\Big\}
\end{multline}
Here $f_{q,g/p}$ are the gluon distribution functions in the proton and
$D_{h/q,g}$ the fragmentation functions of quarks and gluons into the hadron
$h$. The Wilson line correlators $\mathcal{N}_\textrm{F,A}$ 
(as in \eq\nr{eq:wlinecorr}) in the fundamental and adjoint representations
have to be evaluated in at the rapidity scale ($x$ the nuclear wavefunction
corresponding to the produced hadron. A similar approximation for only 
the gluonic contribution can be derived from the $\ktt$-factorized 
formulation discussed in \se\ref{sec:bulk} and has also often been used 
in applications to the spectrum in pA collisions.
A comparison of \eq\nr{eq:pA} using a certain parametrization 
(``DHJ''\cite{Dumitru:2005kb}) with experimental
data is shown in \fig\ref{fig:pA}.
Computing the spectrum in pA collisions at different rapidities
enables one to almost directly see the effects of high energy
evolution in the nuclear wavefunction on the Wilson line correlators
$\mathcal{N}_\textrm{F,A}$. At the starting scale of the evolution
one generically expects a ``Cronin'' enhancement in pA compared to pp 
collisions. This, as can explicitly be seen e.g. in the M.V. model,
is due to the saturation in the nuclear wavefunction that suppresses the
gluon distribution below $\qsa$ and causes an enhancement above
$\qs$ (the ``Cronin peak''). The effect of high energy evolution 
is to wash this peak 
away\cite{Kharzeev:2002pc,*Albacete:2003iq,Kharzeev:2003wz,Blaizot:2004wu,*Blaizot:2004wv}, 
which is seen in the data 
at more forward rapidities, as shown in \fig\ref{fig:rpa}.

\begin{figure}
\resizebox{\textwidth}{!}{
\raisebox{0.5cm}{\includegraphics[height=3cm]{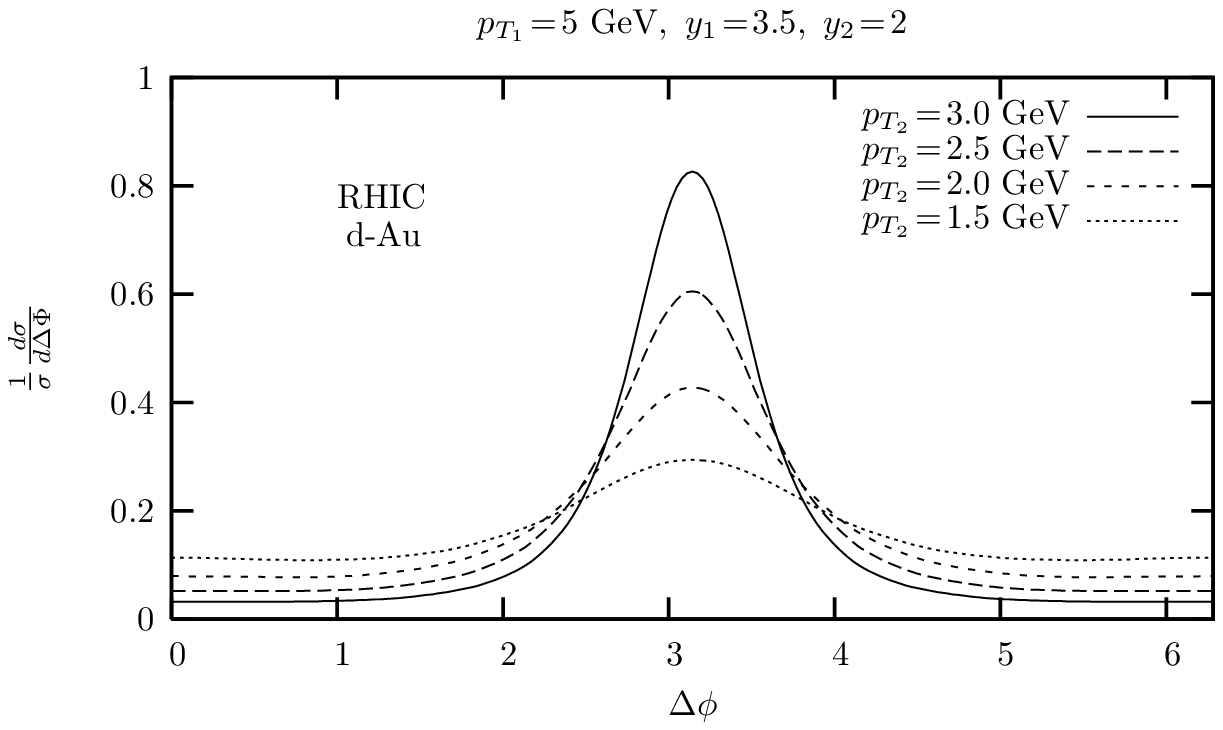}}
\includegraphics[height=4cm]{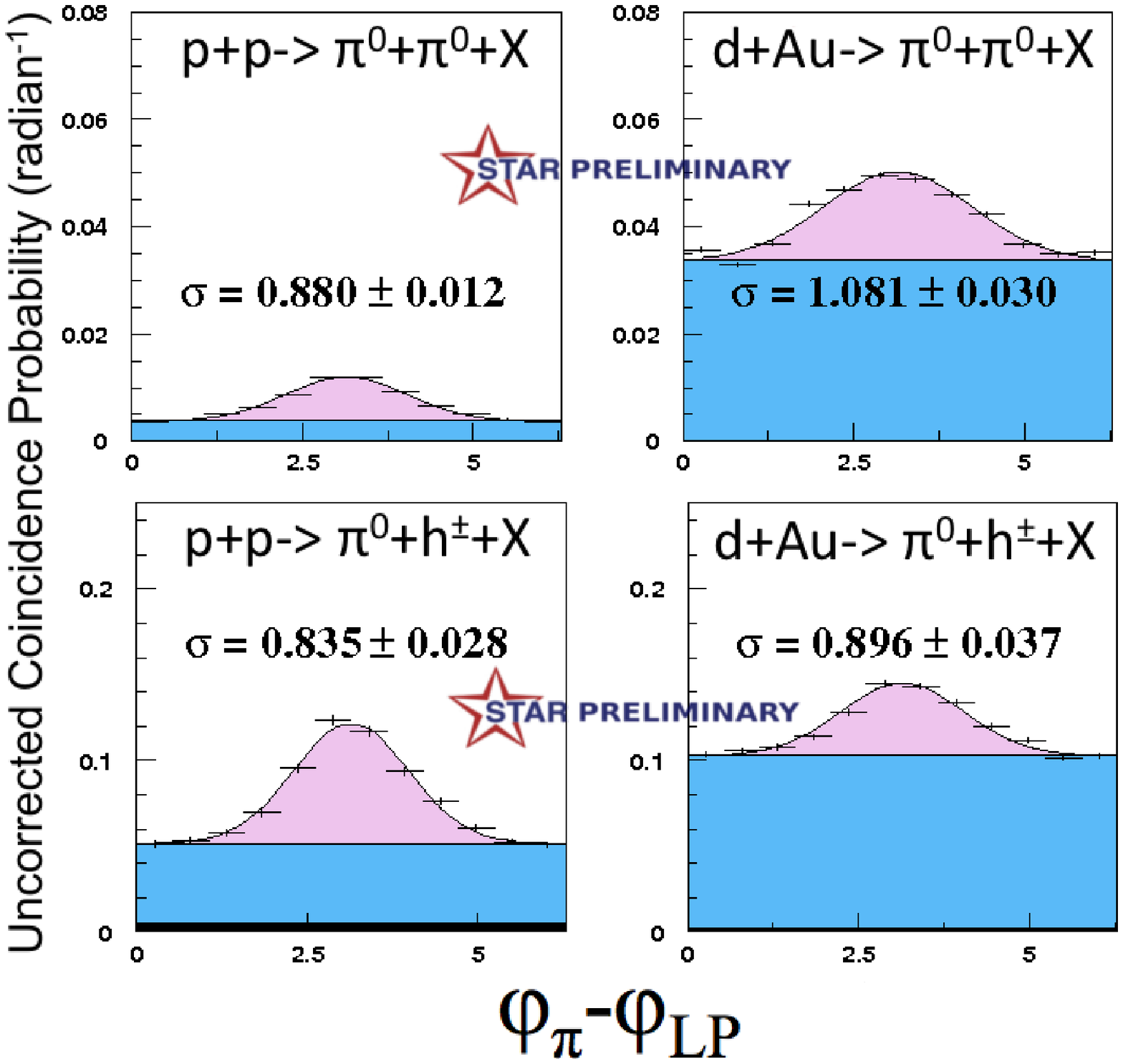}
}
\caption{
Prediction for broadening of the two-particle correlation function in 
deuteron-nucleus collisions
from Ref.\protect\cite{Marquet:2007vb}. 
Right: correspondinf data from the STAR collaboration\protect\cite{Braidot:2009ji}.
}\label{fig:pAcorr}
\end{figure}

A more recent area of activity are two-particle correlations in
dAu collisions\cite{Adams:2006uz,*Adler:2006hi,*Citron:2009eu,Braidot:2009ji}. 
The baseline comparison is the structure observed in 
proton-proton collisions, where high $\ptt$ particles typically originate
from back-to-back jets. The azimuthal structure of the two particle
correlation is thus a rather narrow peak on the away-side resulting from the
fragmentation of the original jet into a narrow cone. The expectation
from the CGC picture is essentially that of an increased collectivity 
in the correlation, signaling itself as a flattening and broadening of the 
away-side peak\cite{Kharzeev:2004bw,*Marquet:2004xa,*Baier:2005dv,*JalilianMarian:2008iz,*Tuchin:2009nf,Marquet:2007vb}. This feature should become more prominent as the 
kinematics of the two particles move towards small $x_A$ (when the saturation
scale becomes larger) and should be very sensitive to the relation
between the trigger $\ptt$'s and $\qs$. A result of a
particular calculation\cite{Marquet:2007vb} is shown in 
\fig\ref{fig:pAcorr}, together from recent experimental data from
STAR\cite{Braidot:2009ji}.

\subsection{Dipole cross section parametrizations}

By now there is a wide variety of saturation-based 
parametrizations of the dipole cross section, mostly fitted to HERA
data. While the aim here is not to give a comprehensive listing, 
let us mention a few. One of the problems of practical parametrizations
is that calculations of DIS observables (as in \eq\nr{eq:f2}) 
use the dipole cross section
as a function of $\rt$, the size of the dipole, what one needs to
compute particle production in pA collisions (as in \eq\nr{eq:pA})
is the Fourier-transform. Although it is not strictly speaking
formally required, one would prefer to have a parametrization
that is positive definite in both position and momentum space,
while retaining some other general features required in different
limits.  It has turned out to be 
a surprisingly difficult mathematical problem\cite{Giraud:2005vx} to 
find such a parametrization. The result of difficulty has 
been that many authors prefer to use different parametrizations
for DIS and for pA studies, thus reducing the
predictive power of the respective calculations.

The early and widely used 
``GBW''\cite{Golec-Biernat:1998js,Golec-Biernat:1999qd} 
dipole cross section is a Gaussian in $\rt$, and 
consequently also in $\kt$:
\begin{equation}\label{eq:gbw}
 \sigma_\textrm{dip}(x,Q^2,\rt) = \sigma_0 (1-e^{-\rt^2 \qs^2(x)/4}),
\end{equation}
with  $\qs^2(x)\sim x^{-\lambda}$ with $\lambda \approx 0.3$ resulting
from the fit to HERA data.
This formula features the idea of parton saturation very clearly. The exponential 
decrease at high momenta is, however, not very physical. At very large 
$\kt$ one would expect $\sigma_\textrm{dip}(\kt)$ to behave as 
$\sim 1/\kt^4$ in order to recover a DGLAP-like
increase of the gluon distribution at high momenta: 
$xG(x,Q^2) \sim \ln Q^2$; one of the attractive features of the MV model 
is precisely that it results in this behavior. At somewhat smaller momenta,
$\ktt \gtrsim \qs$ BK/JIMWLK evolution predicts a different power law
$\sigma_\textrm{dip}(\kt) \sim \ktt^{-2(1+\gamma)}$
with an anomalous dimension $\gamma \approx 0.6$.
In the ``BKW'' parametrization\cite{Bartels:2002cj} the large $\ktt$
behavior is modified by replacing
the exponent $\rt^2\qs^2$ by $\rt^2 xG(x, 1/\rt^2)$, where $ xG(x, 1/\rt^2)$
is a DGLAP-evolved gluon distribution evaluated at the
scale $\mu_0^2 + 4/\rt^2$. This brings the high-$\ktt$ limit
closer to a DGLAP-like behavior, but applying the DGLAP-evolved 
distribution around the saturation scale is not without problems; in
particular it turns out that the best fit to data  is given by
a gluon distribution which \emph{decreases} towards smaller $x$ at the initial
scale $\mu_0^2$. A dipole cross section that reproduces, instead of DGLAP,
the behavior of BK evolution in the extended scaling region $\ktt \gtrsim \qs$
is given by the ``IIM''\cite{Iancu:2003ge} parametrization.

Another simplification in \eq\nr{eq:gbw} is the 
dependence on the impact parameter $\bt$.
 The impact parameter profile  of the dipole scattering amplitude
 can be directly measured from the $t$-distribution of
diffractive events; experiments are consistent with a Gaussian 
in $\bt$ (implying an exponential dependence on $t$).
In \eq\nr{eq:gbw} the impact parameter dependence $T(\bt)$ 
has been factorized from the $\rt$-dependence, which reduced it  
in the total cross section into a constant proton  
area $\sigma_0 = \int \ud^2\bt T(\bt)$. The problematic aspect
of this factorization is that
the unitarity limit of the scattering amplitude is only approached
at $b=0$ (see more detailed discussion in Ref.\cite{Kowalski:2008sa}).
Two parametrizations attempting to treat the $b$-dependence more
consistently, either in a DGLAP-improved approach or parametrizing features
of BK evolution are provided by the IPsat (also ``KT''\cite{Kowalski:2003hm} )
and bCGC models\cite{Kowalski:2006hc}. In these parametrizations the 
saturation scale is impact parameter dependent.

The ``KKT''\cite{Kharzeev:2004yx} and ``DHJ''\cite{Dumitru:2005kb}
parametrizations, also including a BK-like anomalous dimension,
have been used in pA collisions, but not fitted to DIS data.
There are also by now several numerical 
evaluations\cite{GolecBiernat:2001if,*Lublinsky:2001yi,*GolecBiernat:2003ym,*Albacete:2004gw,*Marquet:2005zf,Gelis:2006tb} 
of the BK equations 
numerically both with fixed and running coupling.
The initial condition of the evolution can then be used to fit experimental
data and produce a parametrization\cite{Albacete:2009fh} for use in other 
contexts. A recent application of a direct solution of the BK equation 
to hadron spectra in pA collisions can be found in Ref.\cite{Albacete:2010bs}.

\section{Bulk gluon production in nucleus-nucleus collisions}\label{sec:bulk}

Let us now turn back to the case of the collision of two dense systems.
The first question to understand are
the properties, such as entropy
or energy density, of the initial stage in a heavy ion collision
in the CGC framework. As we have discussed, 
 the initial stage of the collision system
at central rapidities is, to a first approximation, gluonic
and boost invariant. It is therefore far from chemical  and 
thermal equilibrium (in particular anisotropic, with much 
larger transverse than longitudinal momenta).
 The thermalization
of this gluonic system towards a quark-gluon plasma is not
quantitatively understood, in spite of a lot of recent work 
on exploring the instabilities that could dominate this 
stage\cite{Mrowczynski:1994xv,*Mrowczynski:1996vh,*Mrowczynski:2004kv,*Arnold:2003rq,*Romatschke:2003ms,*Romatschke:2004jh,*Nara:2005fr,*Dumitru:2005gp,*Fukushima:2006ax,*Fukushima:2007ja,*Romatschke:2005pm,*Romatschke:2006nk}.

To make a more direct connection to phenomenology one must
thus at this stage 
try to shortcut the thermalization stage by replacing it
with a more schematic relation between the initial gluon spectrum
and the later stage evolution of the system. One simple 
version of this relation, based on experimental observation,
 goes by the name of ``parton-hadron 
duality'', stating that the final state hadron multiplicity
is proportional to that of the initial partons. 
A second way of estimating the relation is based on the
argument that the particle multiplicity cannot decrease
during the thermalization process, due to the second law of 
thermodynamics and the fact that the entropy is essentially 
proportional to the particle density. On the other hand the
energy per unit of rapidity can only decrease (due to $p \ud V$
work, i.e. flow of energy towards the fragmentation region)
during the time evolution.
Thus one can to some extent estimate the relation between the 
initial gluon multiplicity and energy density 
and the final (charged hadron) multiplicity and transverse energy.
In particular it seems reasonable to assume that the dependences
of both on centrality, collision energy and rapidity would be
similar.

\subsection{Calculating the initial gluon multiplicity}

It is thus of direct phenomenological interest to try to calculate 
the initial gluon multiplicity in the CGC framework with, as much
as possible, input values of the parameters determined from 
independent observables, in particular from DIS or pA data.
As already mentioned in \se\ref{sec:glasma}, there are two major 
ways of doing this. One is to directly solve the classical 
Yang-Mills equations numerically (``CYM'', the other is to use
a $\ktt$-factorized approximation (``KLN'') in a situation where it
strictly speaking is not valid. The  results from 
both approaches for the integrated gluon multiplicity,
if not for the spectrum as a function of momentum, turn
out to be very similar.  This can mostly be 
understood by dimensional analysis. Gluon production in a 
central collision, at midrapidity, is a one scale problem
with $\qs$ as the only relevant dimensionful scale.
The gluon density must therefore be 
$\ud N/\ud y \ud^2 \xt \sim \qs^2$. In a noncentral collision
or away from midrapidity the saturation scales of the two nuclei
are different, but the generic result in both the CYM and
KLN calculations is that 
$\ud N/\ud y \ud^2 \xt \sim \qs^2_\textrm{min}$, where 
$ \qs^2_\textrm{min}$ is the smaller of the two saturation
scales. Thus the dependence of the integrated multiplicity on 
centrality is determined by the impact parameter dependence
of the saturation scale, and the dependence on rapidity and
collision energy by its $x$-dependence.  The next
natural observable to look at would be the energy density, which probes
an integral of the gluon spectrum weighted by an extra power of $\ptt$.
This is more sensitive to the large momentum part of 
the spectrum, which would typically behave as $1/\ptt^4$
(as resulting from the expected $1/\ktt^2$--behavior
of the unintegrated gluon distribution) in both cases.
The actual gluon spectrum
from the CYM and KLN calculations can be very different especially
for $\ktt \lesssim \qs$, but it is difficult
to get an experimental handle on the shape of the spectrum.

Typically in KLN calculations, if the value of $\qs$ is fixed
by an external input, the normalization of the result
is adjusted by hand to a reference point in the data. In particular,
in the $\ktt$--factorized approximation in the cases where
one can derive it from the CGC framework the integrated gluon spectrum is not
IR finite; although the divergence is only logarithmic
instead of a power law due to saturation. In practical 
calculations one must thus add an additional regulating 
prescription with the corresponding adjustment to the 
normalization. This adjustment of the normalization is not
possible in the CYM calculation, where the only parameter one has
is essentially $\qs$, which determines both the number and
the typical transverse momentum of the gluons produced. Thus in 
the CYM framework it has turned out to be more realistic to make
a genuine prediction for the normalization of the spectrum based
only on the value of $\qs$ obtained from HERA data; we will discuss
this estimate below. Due to the relative simplicity of the 
analytic $\ktt$--factorized approximation, on the other
hand, it is easier to incorporate more detailed features 
of the transverse coordinate or rapidity dependence.

Most of the original CYM calculations were done using the MV model,
treating the color charge density $g^2\mu$ as a free parameter
to be adjusted to RHIC data. We shall here follow the approach 
of Ref.\cite{Lappi:2007ku} to take these results and transform them
\emph{a posteriori} to parameter-free RHIC postdictions
using the numerically determined relation between $g^2\mu$
and $\qs$ and values of $\qs$ extracted from HERA fits.

Let us first recall the results of the numerical CYM 
calculations of the gluon spectrum in heavy ion 
collisions~\cite{Krasnitz:1998ns,*Krasnitz:1999wc,*Krasnitz:2000gz,Krasnitz:2001qu,Krasnitz:2002mn,Lappi:2003bi,Krasnitz:2003jw,Lappi:2004sf}.
We shall not discuss the numerical procedure here; a recent 
review is given in \cite{Lappi:2009xa}. In order to compare the results
we must, however, comment on one aspect of the numerical implementation
of the MV model. 
In the numerical implementation of the MV model the
 Wilson lines \nr{eq:pathorder} are 
constructed as
\begin{equation}\label{eq:uprod}
U(\xt)  = \prod_{k=1}^{N_y} \exp\left\{ -i g \frac{\rho_k^{1,2}(\xt)}{\nabt^2 + m^2}\right\},
\end{equation}
where the color charges are Gaussian variables with the variance
\begin{equation}\label{eq:discrsrc}
\left\langle \rho^a_k(\xt) \rho^b_l(\yt) \right\rangle =
 \delta^{ab}  \delta^{kl}  \delta^2(\xt-\yt)
\frac{g^2 \mu_A^2}{N_y}.
\end{equation}
The indices $k,l=1\dots N_y$ represent a discretization of the longitudinal 
direction into $N_y$ small steps; 
the continuum limit corresponding to  \eq\nr{eq:pathorder} is achieved for  
$N_y\to \infty$ at constant $g^2\mu_A$. Some kind of 
infrared regulator is needed in order to invert the Laplacian operator
$\nabt^2$, it can be regularized by the scale $m$ in \eq\nr{eq:uprod}.
As pointed out in Ref. \cite{Lappi:2007ku} (see also\cite{Fukushima:2007ki}), 
the relation between the parameter $g^2\mu$ of the model and
the physical length scale $\qs$, the correlation length of the
Wilson lines in the transverse plane, depends on  $N_y$. 
The conceptual picture CGC framework is that of a Wilson line built up
from infinitesimal steps in rapidity, i.e. $N_y\to \infty$. The 
early numerical implementations, on the other hand, used $N_y=1$
for simplicity. When expressed in terms of the physical length scale
$\qs$ the results depend very little on the way the Wilson lines
were constructed. However, when comparing the results of the numerical
calculations to phenomenology and using values of $\qs$ extracted 
independently one must take care to use the correct relation
between $\qs$ and $g^2\mu$. As shown in\cite{Lappi:2007ku}, the
relation corresponding to $m\to0,N_y=1$; i.e. the prescription
used in the numerical CYM calculations, is $\qs \approx 0.57 g^2\mu$.

With this conversion between $\qs$ and $g^2\mu$ explicitly stated
let us then proceed to the result of the numerical CYM 
computations. 
The energy and multiplicity per unit rapidity 
can be parametrized as
\begin{eqnarray}
\frac{\ud N}{\ud \eta} &=& \frac{(g^2\mu)^2 \pi \ra^2}{g^2}f_N 
\label{eq:fn}
\\
\frac{\ud E_T}{\ud \eta} &=& \frac{(g^2\mu)^3 \pi \ra^2}{g^2}f_E.
\end{eqnarray}
The numerical result (see in particular Refs.~\cite{Lappi:2003bi,Krasnitz:2003jw})
is $f_E\approx 0.25$ and $f_N\approx 0.3$. 
We can also recast this expression in terms of the the 
``liberation coefficient'' $c$, 
introduced by A. Mueller~\cite{Mueller:1999fp,Kovchegov:2000hz,Mueller:2002kw},
The liberation coefficient
is defined by writing the produced gluon multiplicity as
\begin{equation} \label{eq:libc}
\frac{\ud N}{\ud^2 \xt \ud y} = c \frac{\cf \qs^2}{2 \pi^2 \as}.
\end{equation}
With \eq\nr{eq:fn} this leads to
\begin{equation} \label{eq:libc2}
c = \frac{\pi f_N}{2 \cf} \left( \frac{g^2\mu}{\qs} \right)^2.
\end{equation}
The original expectation was that $c$ should be of order 
unity\cite{Baier:2002bt,Mueller:2002kw}. The
analytical calculation by Y. Kovchegov~\cite{Kovchegov:2000hz}
gave the estimate  $c \approx 2\ln2 \approx 1.4$. 
With $f_N = 0.2$ in \eq\nr{eq:fn} and $\qs/g^2\mu \approx 0.57$
we see that the  CYM result for the
liberation coefficient is $c \approx 1.1$. A demonstration of the 
independence of this number on the details of the
Wilson line correlator can be made by replacing the MV model distribution by 
another one. For example, if the Wilson lines are constructed 
explicitly to correspond to the IPsat\cite{Kowalski:2003hm,Kowalski:2007rw}
(Kowalski-Teaney) or bCGC (\cite{Iancu:2003ge,Kowalski:2007rw}) 
parametrization of HERA data, the result for the liberation coefficient
$c$ is the same as in the MV model across the whole energy range
from RHIC to the LHC\cite{Lappi:2008eq}. 
This is the result in spite of the very different
forms of the gluon spectra at high $\ktt$ in the models.

The estimates for the values of $\qs$ based on HERA data, nuclear geometry
and attempts to fit the limited available nuclear DIS 
data\cite{Freund:2002ux,Armesto:2004ud} vary somewhat, especially 
due to the degeneracy between the proton size and the value of $\qs$ 
in the HERA fits; see e.g.\cite{Lappi:2007ku}
for a comparison of the numerical values. We take here the IPsat 
estimate\cite{Kowalski:2007rw} that removes this interplay by fixing
the size of the proton from diffractive data, where it can be directly 
measured. The estimate amounts to  $\qs \approx 1.2 \gev$ 
for an average central RHIC collision at 
midrapidity\cite{Kowalski:2007rw,Lappi:2007ku}.
As noted previously, this corresponds to  
the MV model parameter $g^2 \mu \approx 2.1\gev$ (at $N_y=1$)
in the CYM simulations. Plugging this number into \eq\nr{eq:fn}
leads to the estimate of $\frac{\ud N}{\ud y} \approx 1100$
gluons in the initial stage of a central gold-gold collision
at midrapidity at RHIC. This is very close to the observed final 
total (charged+neutral) particle multiplicity. In other words,
recalling our discussion earlier in this section, this points 
towards a very rapid thermalization that practically conserves particle
number, and leaves very little room for higher order effects to
increase the multiplicity during the thermalization stage.

\subsection{Energy and rapidity dependence}

\begin{figure}
\resizebox{\textwidth}{!}{
\includegraphics[height=5cm]{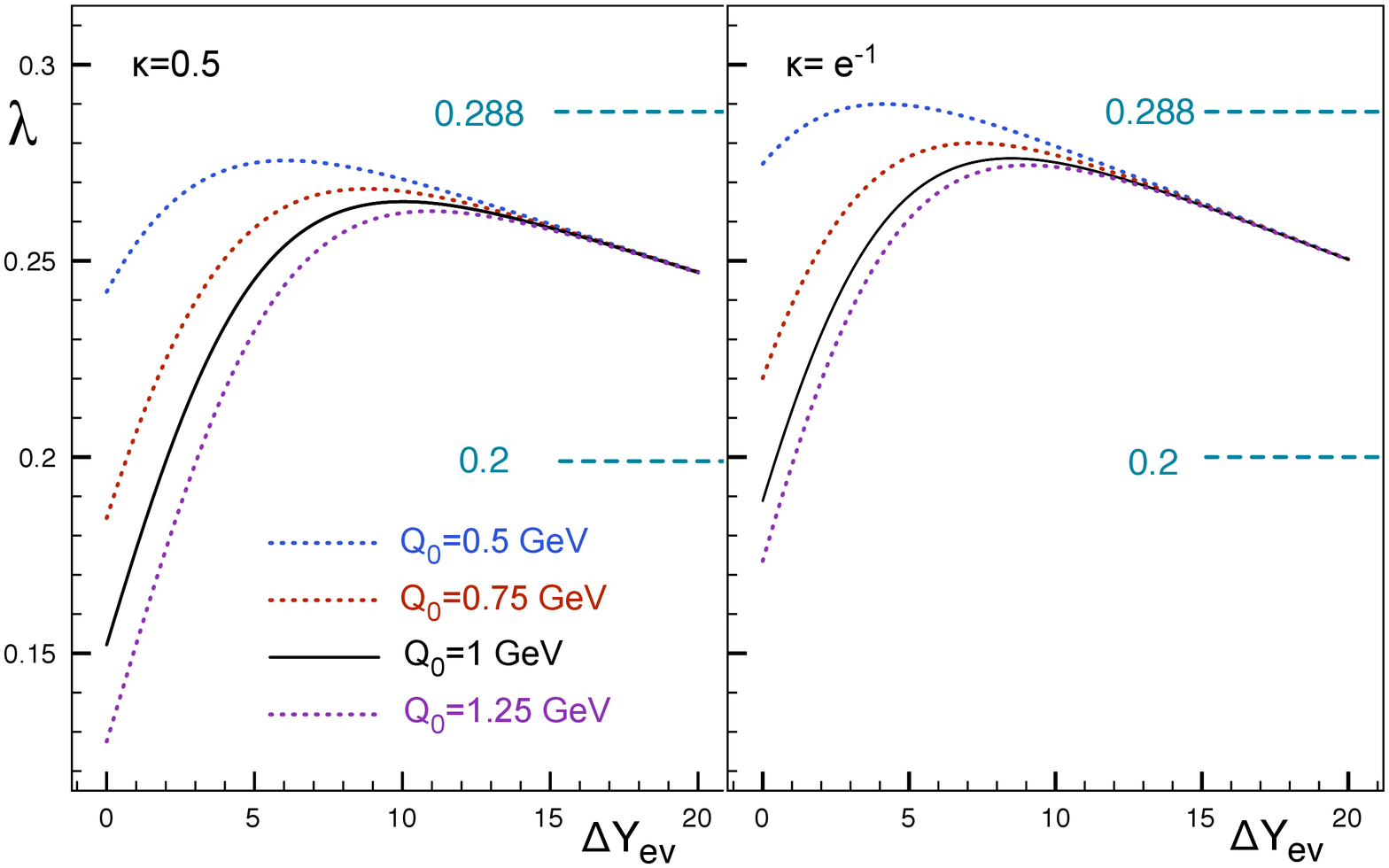}
\includegraphics[height=5cm]{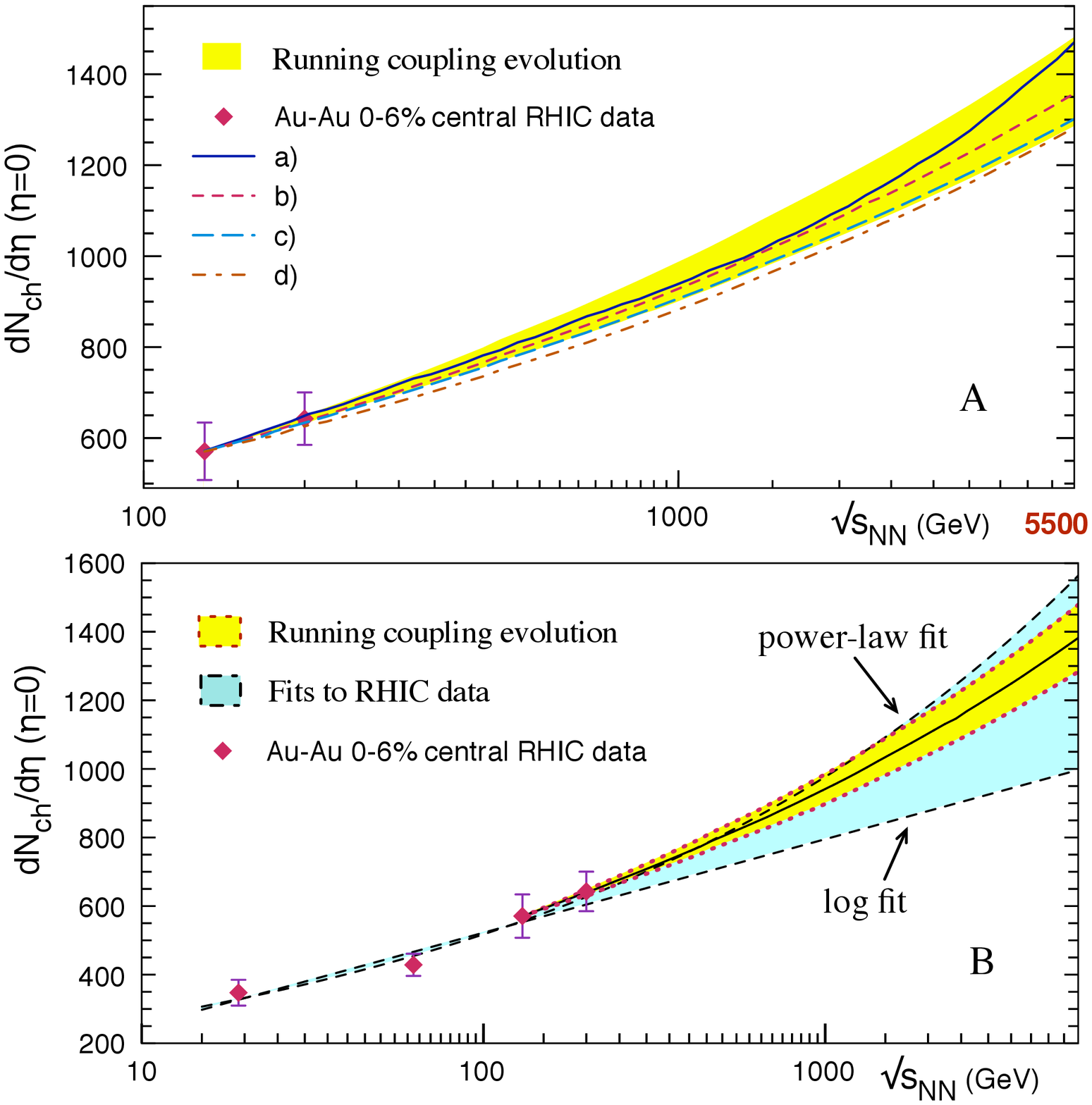}
}
\caption{
Predictions for the evolution speed from  the running coupling BK equation
and  the subsequent gluon multiplicity extrapolated to LHC energies from 
Ref.\protect\cite{Albacete:2007sm}.
}\label{fig:multi}
\end{figure}

As we discussed, the gluon multiplicity is, across different 
parametrizations to a very
good approximation proportional to $\pi \ra^2 \qs^2/\as$. Thus the
predictions for LHC collisions depend mostly on the energy dependence
of $\qs$. On this front there is perhaps more uncertainty than is
generally acknowledged, the estimates for $\lambda= \ud \ln \qs^2 /
\ud \ln 1/x$ varying between $\lambda=
0.29$~\cite{Golec-Biernat:1998js} and $\lambda=
0.18$~\cite{Kowalski:2006hc} in fixed coupling fits to HERA data, with
a running coupling solution of the BK equation giving something in
between these values~\cite{Albacete:2007sm}. This dominates the
uncertainty in predictions for the LHC multiplicity (see
\fig\ref{fig:multi}).

The RHIC collision energy is still too slow to clearly see any
saturation effects in the rapidity dependence of the multiplicity
around $y=0$. A simple estimate for the effects of large $x$ physics,
such as momentum conservation, is to consider the typical
$(1-x)^4$-dependence of gluon distributions at large $x$.  Inserting
$x=e^{\pm y} \langle p_\perp \rangle / \sqrt{s} $ leads to the
estimate $\Delta y \sim \sqrt{8 \sqrt{s}/ \langle p_\perp \rangle}$
for the rapidity scale at which the large $x$ effects contribute to
the rapidity distribution around $y=0$. 
This means that the $(1-x)^4$-behavior starts to dominate
the rapidity dependence of the multiplicity around midrapidity
at a scale
of $\Delta y \sim 4$ RHIC and $\Delta y \sim 19 $ at the LHC.
Note also that the large $x$ contribution is an
effect of order 1 at this scale, whereas small $x$ evolution can be
expected to give a much smaller effect~\cite{Lappi:2004sf} at a
rapidity scale $\Delta y \sim 1/\as \sim 3$. Only at the LHC the large
$x$ effects will be mostly absent around midrapidity and one has a
good possibility of seeing a clear signal of
CGC effects in the rapidity dependence of  the multiplicity.

\section{Transverse geometry}\label{sec:transverse}

Relativistic heavy ion collisions can take place with any impact parameter,
from very peripheral ones that should look like simple nucleon-nucleon 
collisions to central ones that actually produce a system of the size of the 
colliding nuclei. Experimentally this provides a tool to study quantities
as a function of the system size, assuming that one is able to detect
the centrality. The impact parameter not being  a measurable
quantity \emph{per se}, there are basically two methods used
to extract the collision centrality in experiments. One is to measure the 
noninteracting spectator nucleons in the zero-degree calorimeters (ZDC) and the 
other to make a simple assumption on the dependence of the charged 
multiplicity in some part of the detector on the size of the system.
This is usually done via a Monte Carlo Glauber calculation, which enables one to
compute 
the distributions of $\npart$ and $\ncoll$ (the numbers of participant nucleons
and binary nucleon-nucleon collisions) for a fixed impact parameter; assuming
that the individual nucleon-nucleon collisions are independent of each other.
Using a simple ansatz for the dependence of the charged particle multiplicity
on $\npart$ and $\ncoll$ one can then
divide the events of the whole minimum bias data set  into 
centrality classes and estimate the typical impact parameters corresponding
to each class
(see \cite{Miller:2007ri} for a review of Monte Carlo Glauber modeling).

Besides the ZDC data there are few ways to independently check the consistency
of the MC Glauber framework for understanding the collision geometry. One is
is nevertheless comforted by a general impression
that the picture seems to work very consistently, depending
weakly on the details of the MC Glauber model and successfully parametrizing
a wide range of phenomena. 
In particular it turns out that the bulk particle production mechanism
is such that the charged particle multiplicity is roughly proportional 
to the number of participant (``wounded'') nucleons, with a constant of 
proportionality increasing slightly as one goes to more central collisions.
A MC Glauber computation is, however, merely a parametrization of data: it does
not contain any dynamics and therefore does not provide a microscopical 
explanation of the particle production mechanism.

\subsection{Centrality dependence of multiplicity}

\begin{figure}
\resizebox{\textwidth}{!}{
\includegraphics[height=5cm]{multiwphobos.eps}
\includegraphics[height=5cm]{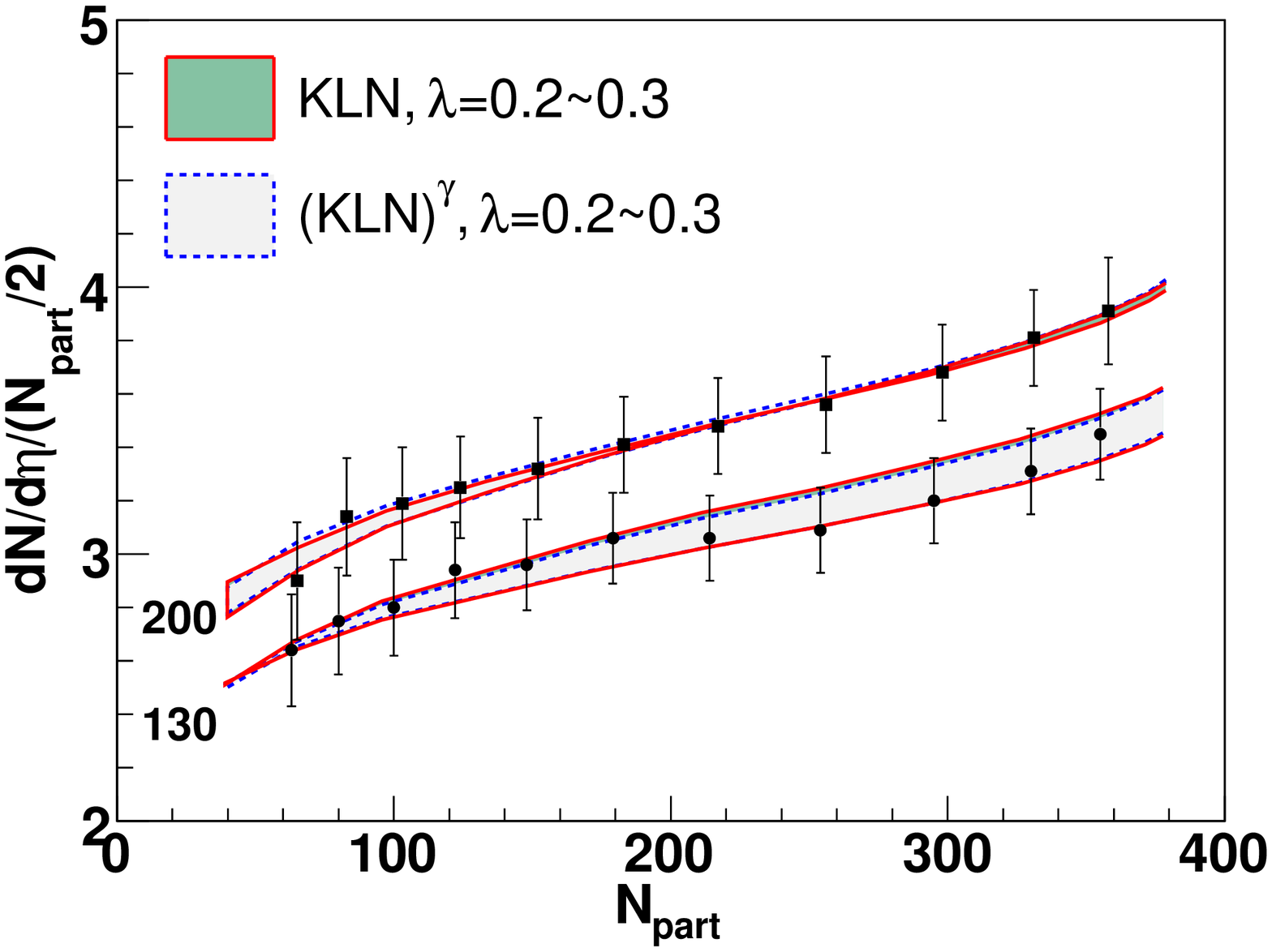}
}
\caption{
Centrality dependence of the multiplicity. Left: 
CYM calculation~\protect\cite{Lappi:2006xc}. Right:
KLN calculation~\protect\cite{Drescher:2006ca}.
}\label{fig:multinpart}
\end{figure}

The approximate proportionality of $\ud N/\ud \eta$ to 
$\npart$ is a natural result in the CGC framework. The saturation scale $\qs$ being
the only dimensionful parameter in the problem the multiplicity per unit area
should be $\sim 1/\qs^2$. In a fully central collision the saturation scales
of the two nuclei at a transverse coordinate $\xt$ are equal to each other and 
proportional to the nuclear thickness (and therefore to $\npart$) at the point 
$\xt$; integrating over the transverse plane gives a multiplicity proportional to
$\npart$. 

For noncentral collisions the situation at a fixed transverse coordinate $\xt$ is 
asymmetric since the two saturation scales $\qsa$ and
$\qsb$ are not equal to each other. In this case 
the multiplicity can also have a dependence on $\qsa/\qsb$. 
Parametrically, the  spectrum of produced gluons in this case  
behaves~\cite{Dumitru:2001ux,Blaizot:2004wu} as 
\begin{eqnarray} 
\frac{\ud N}{\ud^2 \xt \ud^2 \pt} & \sim & \ln(p_T), \quad p_T < \qs_1 
\\ 
& \sim & \frac{\qs_1^2}{p_T^2}, \quad \qs_1 < p_T < \qs_2 
\\ 
& \sim & 
\frac{\qs_1^2 \qs_2^2}{p_T^4}, \quad p_T > \qs_2 ,
\end{eqnarray} 
where $\qs_1$ and $\qs_2$ are the smaller and larger of the saturation 
scales $\qsa$ and $\qsb$. 
Integrated over transverse momenta, this gives 
\begin{equation} \label{eq:pamulti} 
\frac{\ud N}{\ud^2 \xt} \sim \qs_1^2 
\end{equation} 
neglecting logarithmic corrections 
$\sim \ln\left( \qs_2/\qs_1\right)$. 

Both the CYM and KLN types of 
calculations\cite{Kharzeev:2000ph,Lappi:2006xc,Drescher:2007ax}
 reproduce quite well the  experimental
data on the centrality dependence of the multiplicity, as shown in 
\fig\ref{fig:multinpart}. In the CYM calculation all the deviations 
from a strict proportionality $\ud N/\ud \eta \sim \npart$ 
come from the (parametrically logarithmic) deviations
from the strict proportionality to the smaller one of the saturation
scales, \eq\nr{eq:pamulti}. In the KLN calculations, especially
in the earlier versions where the geometry was treated in a more simplified
way, a significant part of the deviation  from 
$\ud N/\ud \eta \sim \npart$ was argued to result from the running of
the coupling constant in $\ud N/\ud \eta \sim \pi \ra^2\qs^2/\as(\qs)$.
The deviations from a number of participant scaling in the experimental
data are small enough that they do not allow one to discriminate between the
relative importance of these effects.

\subsection{Eccentricity of the initial state}\label{subsec:ecc}

\begin{figure}
\resizebox{\textwidth}{!}{
\includegraphics[height=5cm]{nucl-th0609021ecc.eps}
\includegraphics[height=5cm]{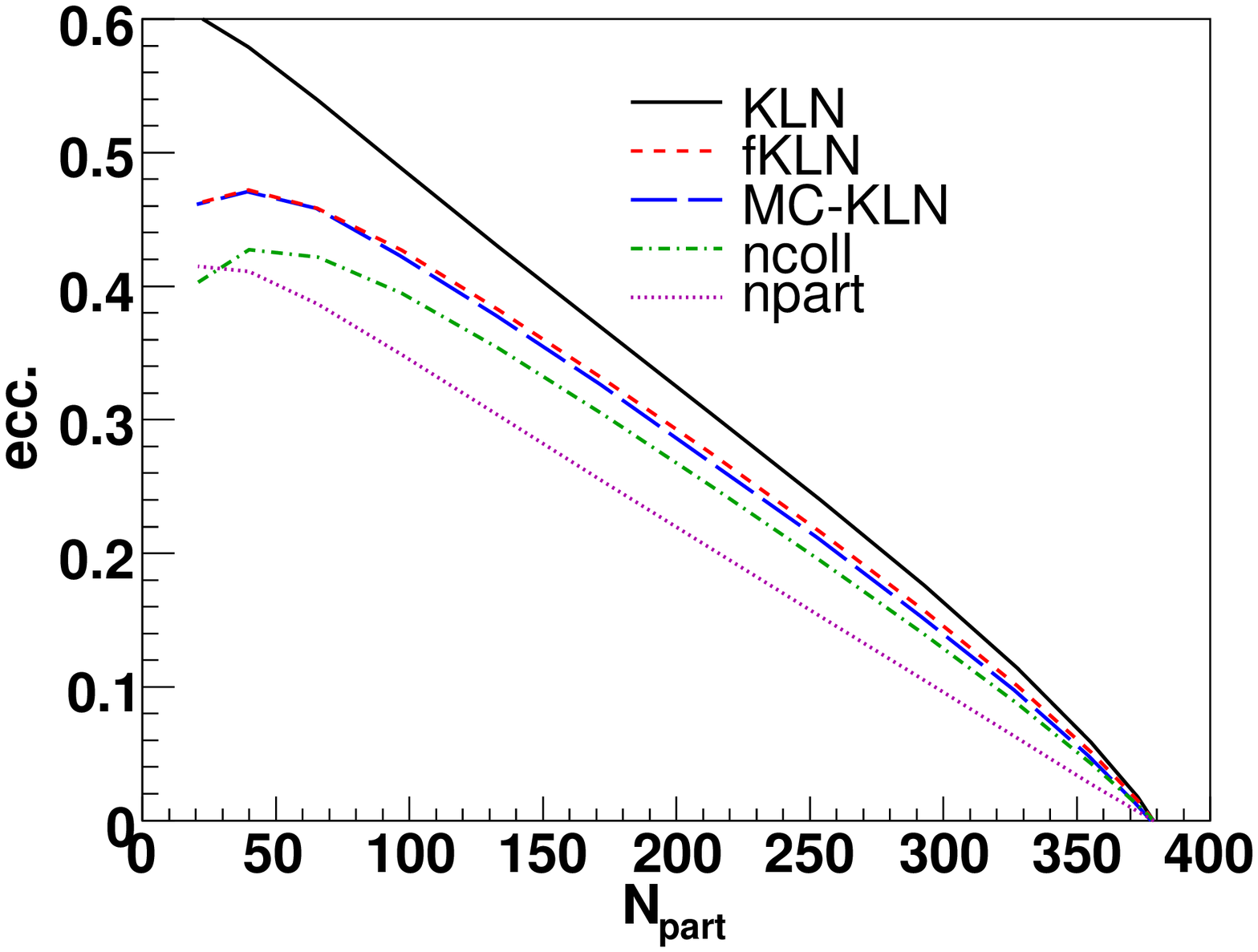}
}
\caption{
Left:
The original KLN eccentricity~\protect\cite{Hirano:2005xf} compared to the
CYM result~\protect\cite{Lappi:2006xc}. 
Right: A subsequent  revision of the KLN model was to reformulate the nonuniversal
saturation scale in terms of fluctuations in the participant 
nucleons led to a lower eccentricity estimate 
(``fKLN'')~\protect\cite{Drescher:2006ca} practically in agreement with the CYM 
result and very close to binary collisions scaling.
} \label{fig:ecc}
\end{figure}

A striking signal of collective behavior of the matter
produced at RHIC is elliptic flow (for a recent review, see\cite{Sorensen:2009cz}).
 Comparing hydrodynamical calculations with flow
is a way to address fundamental properties of the medium, such as the viscosity,
but this comparison requires understanding of the initial conditions of the hydrodynamical
evolution, particularly the initial eccentricity for elliptic flow.
The original general consensus some years ago 
 was that ideal hydrodynamics  is in good agreement
with the experimental data.  Among other caveats this general conclusion
also supposes that one is relatively free to choose the initial 
conditions---in particular the transverse geometry---of 
the hydrodynamical calculation to fit the experimental data
(see e.g. the thorough comparison of initial conditions\cite{Kolb:2001qz}).
This claim has had to be reevaluated more recently after 
it was argued (using a KLN-type 
calculation\cite{Hirano:2005xf,Drescher:2006pi,Kuhlman:2006qp}) that CGC
results in a larger initial eccentricity than traditionally used 
in hydrodynamical calculation
(mostly participant or wounded nucleon, often called
``Glauber'' initial conditions).
This leaves more room for viscosity 
in the hydrodynamical evolution. The CGC estimate has now settled to a lower
value than first argued in\cite{Hirano:2005xf} due to further developments
that we will outline below. The result remains, however, that generically 
CGC calculations tend to predict an initial state with an eccentricity 
close to or slightly above $\ncoll$ scaling, which is
larger than often  used in hydrodynamics.
The eccentricity obtained in the CYM calculation is demonstrated
in \fig:\ref{fig:ecc}.
Everything else staying equal, the larger eccentricity of the initial
state can be compensated by the effect of a larger viscosity in order to produce
the observed elliptic flow $v_2$. When other parameters in the fit are also allowed
to change, the situation can become more complicated and, in some cases, contrary
to this basic intuition. For example in the ``Knudsen number'' fit
of\cite{Drescher:2007cd} assuming a finite cross section (and thus deviations
from ideal hydrodynamics including viscosity) also the equation of state 
was allowed to vary 
as a fit parameter. The result of the fit was that the CGC initial conditions
actually corresponded to a smaller viscosity (larger cross section); the
observed smaller $v_2$ being achieved by a softer equation of state
($c_\textrm{s}= 0.22$ vs $0.3$ for the Glauber initial conditions).

\begin{figure}
\includegraphics[width=0.45\textwidth]{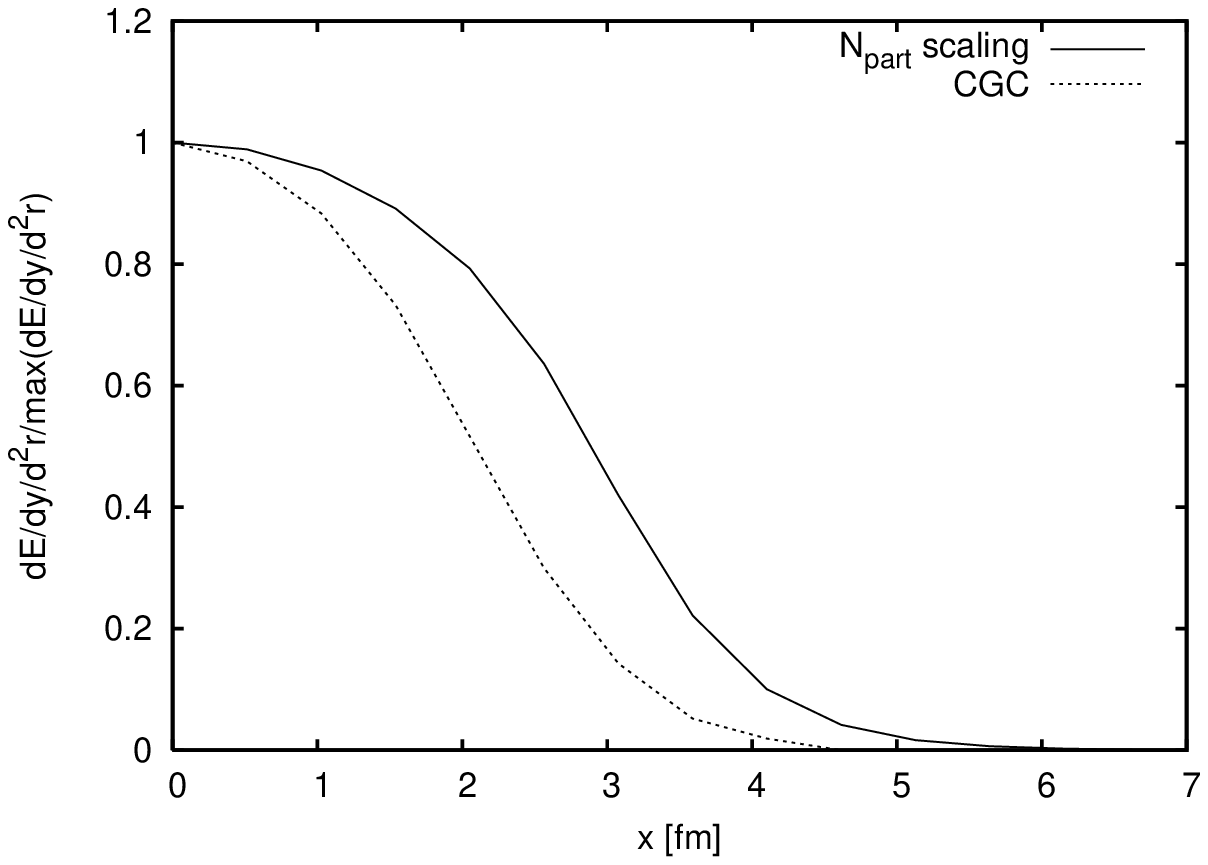}
\hfill
\includegraphics[width=0.45\textwidth]{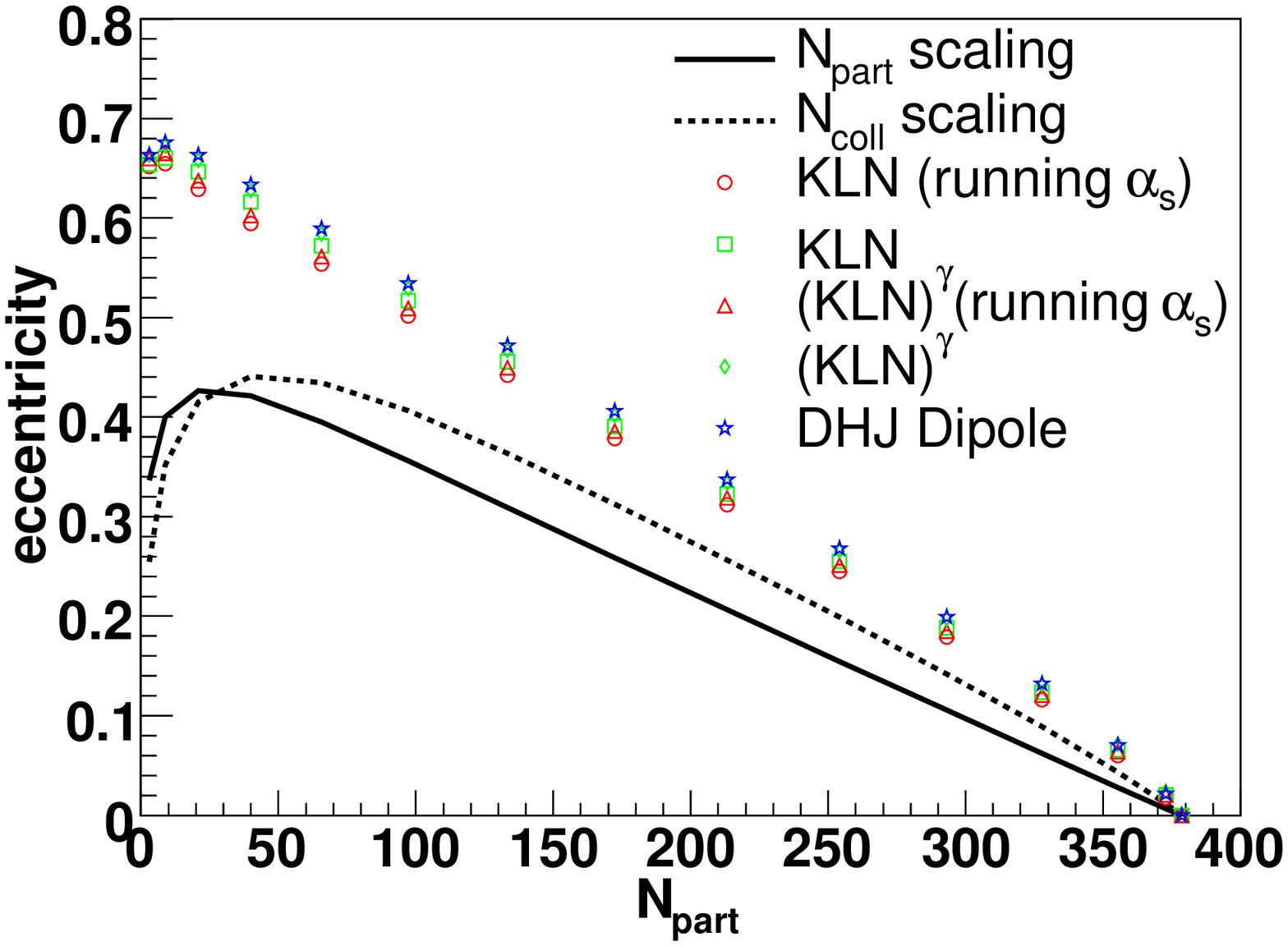}
\caption{
Left: energy density as a function of the transverse coordinate 
in the original KLN definition compared to participant number scaling. Right: 
the KLN eccentricity for different unintegrated gluon distribution parametrizations.
Plots from Ref.\protect\cite{Drescher:2006pi}.
}\label{fig:klnecc}
\end{figure}

As we saw previously, in the case of two different saturation scales
at a point  in the transverse plane, the gluon multiplicity only depends
on the smaller one of the two. The initial eccentricity, on 
the other hand, is computed from the energy density\footnote{
In particular in the CYM calculation, where one is explicitly faced
with the fact that the energy density is practically the only local 
gauge invariant observable available. The gluon spectrum is determined
by Fourier-transforming the field modes and is therefore necessarily
nonlocal at a scale $1/\ptt$. In a  (semi)analytical calculation such 
as KLN one can formally evade this problem.}, which behaves 
parametrically as $\qs_1^2 \qs_2$, where again $\qs_1 < \qs_2$.
In the original KLN calculations the transverse coordinate dependence
of the saturation scale was taken as $\qsa^2(\xt) \sim \nparta(\xt)$,
where $\nparta(\xt)$ is the density of participating nucleons in 
nucleus A. This has the advantage of being directly connected to the well
established Glauber modeling of the geometry also used to bin the experimental
data. To understand the effect of this prescription
on the calculated eccentricity one
must study what happens in a noncentral collision. The important
region for this is the one at the edge of one of the nuclei
(nucleus A) and the center of the other one (B). In this case the definition
leads to both saturation scales being small (since $\npartb$ goes
to zero outside of nucleus A). This is conceptually problematic, since
 one would expect the saturation scale to be large in the center of 
nucleus B. The KLN prescription is therefore nonuniversal: the
value of $\qsb$ is not a property of nucleus B alone, but is
determined by a final state effect; namely the presence or not
of the other nucleus A at a given transverse coordinate. Since
$\varepsilon \sim \qsb \qsa^2$ in this region, this prescription leads
to a suppression in the energy density in the edge of the interaction
region and a larger eccentricity, basically independently
of the precise form of the unintegrated gluons distribution 
(see \fig\ref{fig:klnecc}).
To remedy this nonniversality problem in the KLN model 
it has subsequently been reformulated into what is known as
 the  fKLN\cite{Drescher:2006ca} (fluctuating KLN) model, where
one first introduces explicitly the nucleon-nucleon cross section to 
determine whether a given nucleon participates in the scattering. After
this step the resulting gluon spectrum is the calculated with universal 
saturation scales in the two nuclei. The fKLN model, has practically by 
construction, the property that a) the saturation scales are universal,
b) the original KLN centrality dependence of the multiplicity is reproduced
(due to the property \nr{eq:pamulti}) and c) it reproduces event-by-event 
fluctuations\cite{Drescher:2007ax} that are close to ones given by Monte
Carlo Glauber calculations. (For experimental results on 
$v_2$ fluctuations see e.g. Ref.\cite{Sorensen:2008zk,*Alver:2008hu}).
The resulting eccentricity,
as shown in \fig\ref{fig:ecc} (right) becomes smaller, 
closer to the result (similar to $\ncoll$ scaling)
obtained in the straightforward and universal CYM calculation.
The price to pay for this modification is a nonuniversality of another
kind: namely the $\ktt$-factorized formalism becomes different for 
AA-collisions than it is for the case of proton-proton collisions.
In the treatment of AA collisions in the fKLN formulation there
is an additional step of computing first a collision probability
of individual constituents of the colliding objects (nucleons in 
a nucleus) which is not invoked in the pp-case. In other words, nuclei are,
even at high energy and high $\ptt$, not treated as consisting of
quarks and gluons, but as fundamentally more complicated
objects to which the usual (collinear) QCD factorization theorems 
do not apply. One curious consequence of this feature is that
the saturation scale of the nucleus in the fKLN model approaches
a constant, not zero, at arbitrarily large distances
outside the nucleus.

\section{Correlations in the glasma}\label{sec:corr}

\begin{figure}
\resizebox{\textwidth}{!}{
  \includegraphics[height=5cm]{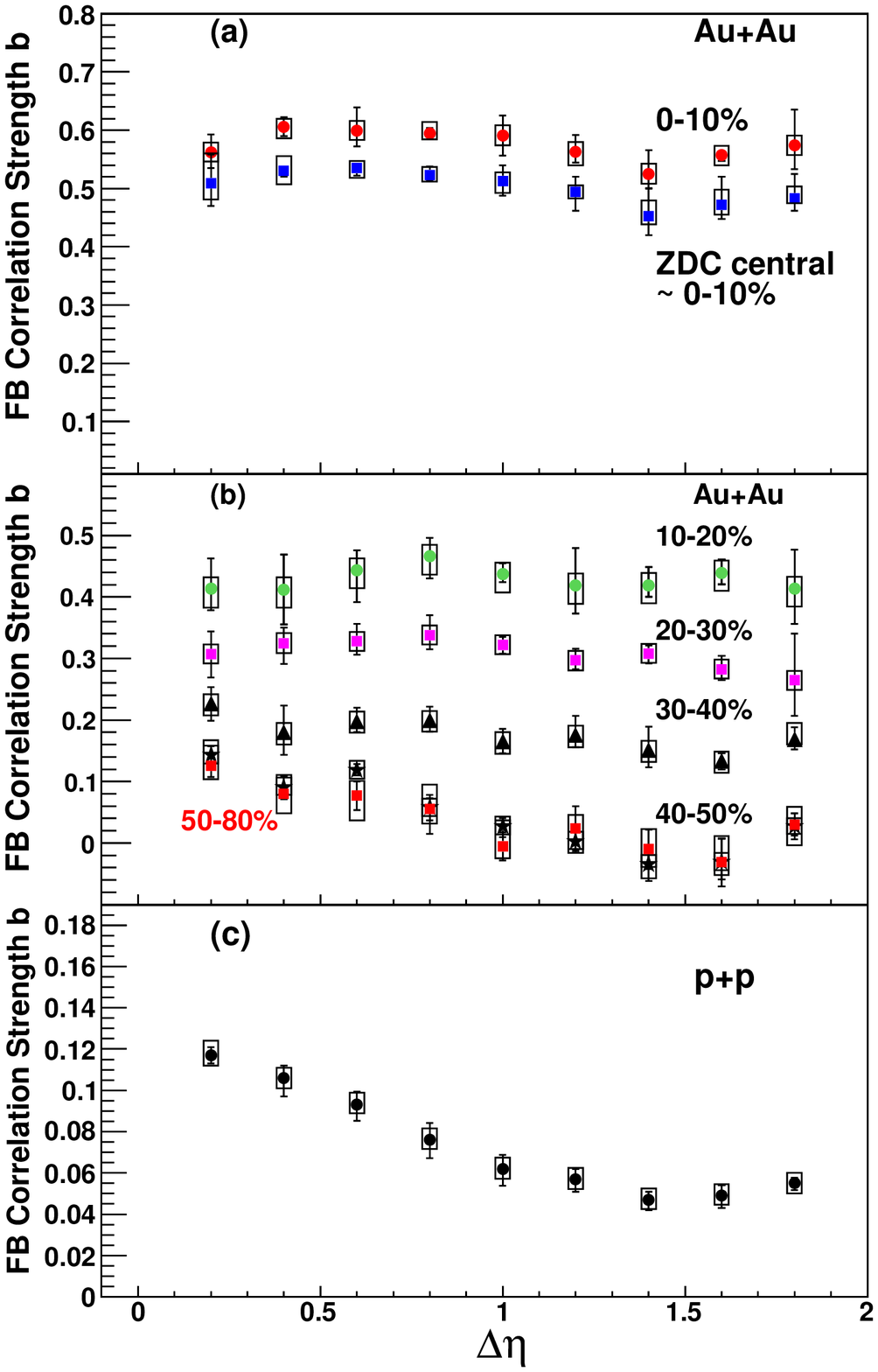}
  \resizebox{!}{5cm}{
    \begin{minipage}[b]{5cm}
      \noindent
      \includegraphics[width=5cm]{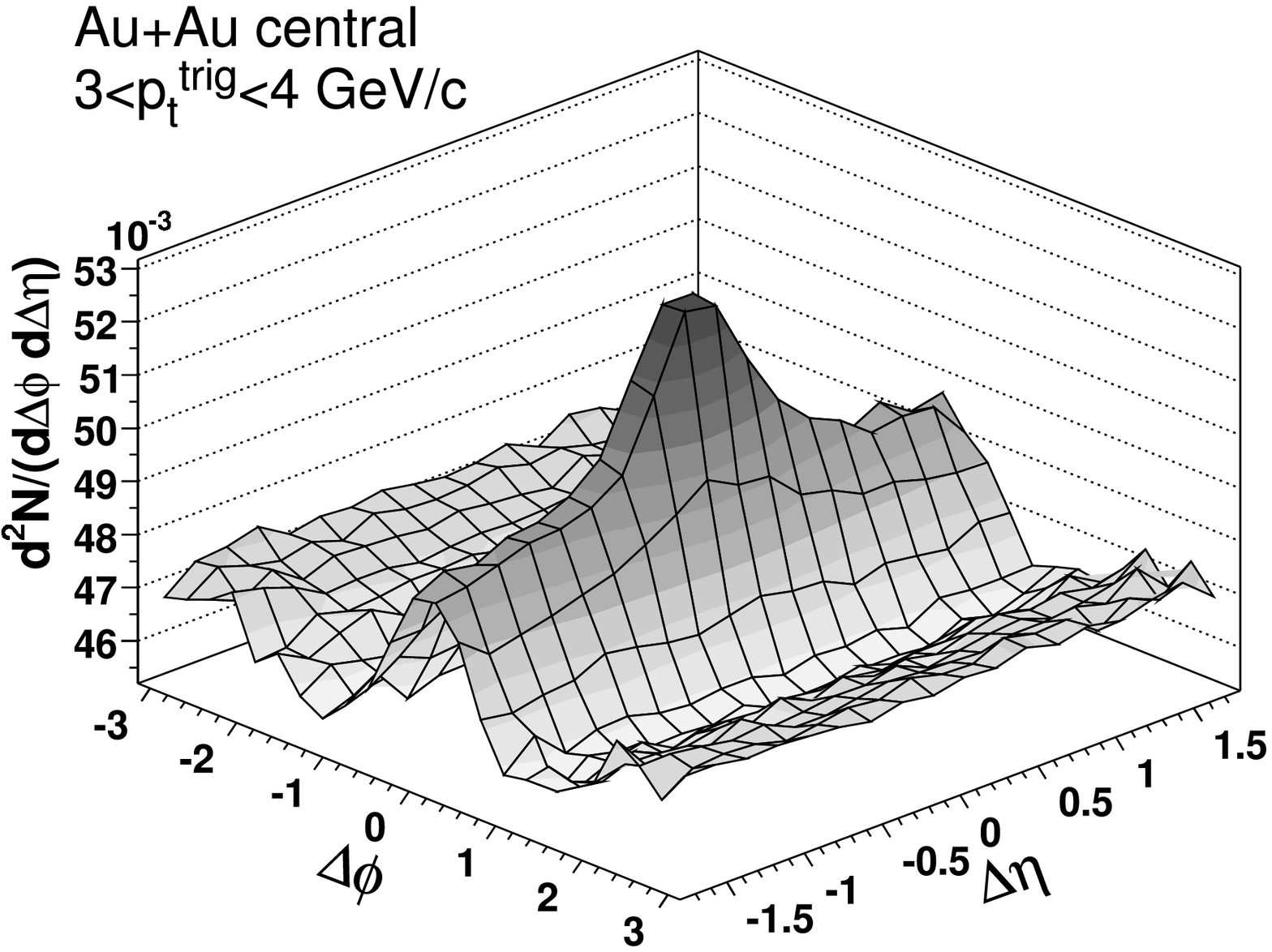}
      \includegraphics[width=5cm]{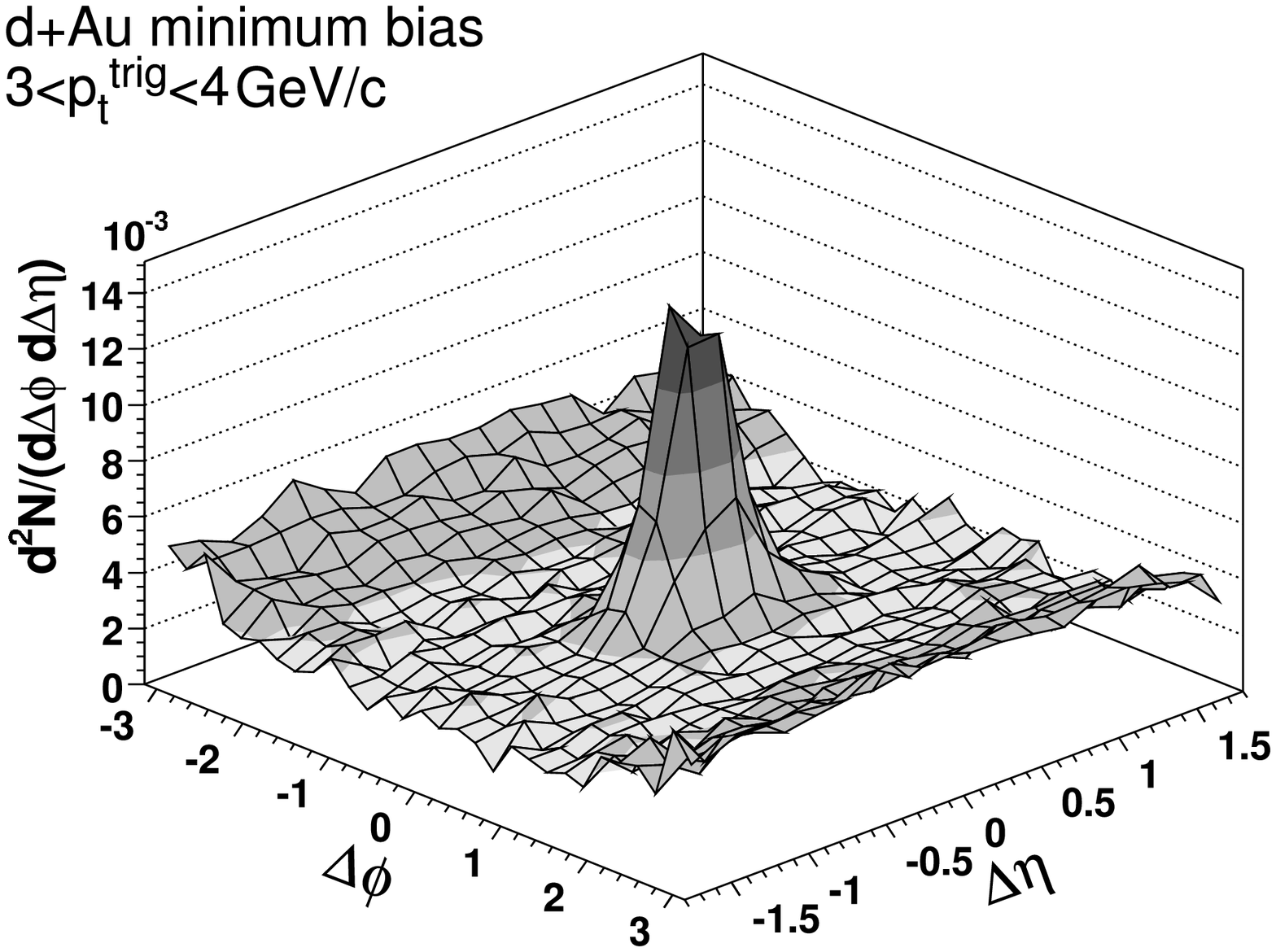}
    \end{minipage}
  }
}
\caption{
Left: forward-backward multiplicity correlations from 
STAR\protect\cite{Abelev:2009dq}
Right:
Experimental 2-particle correlation data\protect\cite{Abelev:2009qa}
in AuAu and dAu collisions, showing the elongated ridge structure in the
former case.
}\label{fig:expcorrs}
\end{figure}

So far we have discussed how the CGC picture of the small $x$ nuclear
wavefunction can be used to compute bulk properties of the
initial state of a heavy ion collision. Experiments do not probe this
initial state directly, because the system goes through a complicated
time evolution before the hadronization stage. A good candidate for a more
direct experimental probe of the glasma stage is provided by different kinds of
correlation measurements, and they have recently been a focus of 
both experimental and theoretical activity. 
Particles that are produced far in rapidity can, by causality,
only be correlated at early times, and should therefore be
directly related to the properties of the glasma\footnote{Although 
when making precise estimates
one must be more careful and distinguish spacetime, momentum space and pseudorapidity
and understand shorter range correlations such as those from hadronization.}.
Examples of these long range correlations are
the ``Ridge'' structure of two particle 
correlations in central heavy ion 
collisions~\cite{Adams:2004pa,*Daugherity:2008su,Putschke:2007mi,*Wenger:2008ts,*Alver:2008gk,*Nagle:2009wr,Abelev:2009qa}
and of long range rapidity correlations in 
particle multiplicities~\cite{Abelev:2009dq,Tarnowsky:2008am}
(see \fig\ref{fig:expcorrs})\footnote{
The interpretation of this multiplicity correlation data is somewhat puzzling  
because of  the interplay with the impact parameter induced, 
purely geometrical, correlations~\protect\cite{Lappi:2009vb}.}.
Another interesting correlation measurement are electric charge-reaction plane 
correlations\cite{Voloshin:2008jx,*Voloshin:2009hr,Abelev:2009uh,*Abelev:2009tx}
that can be due to CP-violating fluctuations in the initial stage.
As we shall argue, these correlations also arise naturally in the glasma.
Correlations between the electric charge and the reaction plane 
are most likely mediated by the magnetic field caused by the noncentral collision 
of two positively charged ions; it is hard to think of another quantity that
would as directly couple to both the reaction plane direction and the 
electric charge. Since this magnetic field dies away very fast with $\tau$,
the QCD part if the explanation must be sensitive to the gluon fields at
the earliest stage of the collision.
In the following we shall discuss some features of how these
correlations can be understood in the glasma framework.

\subsection{Multigluon correlations in the boost invariant case}

\begin{figure}
\centerline{\includegraphics[width=0.8\textwidth]{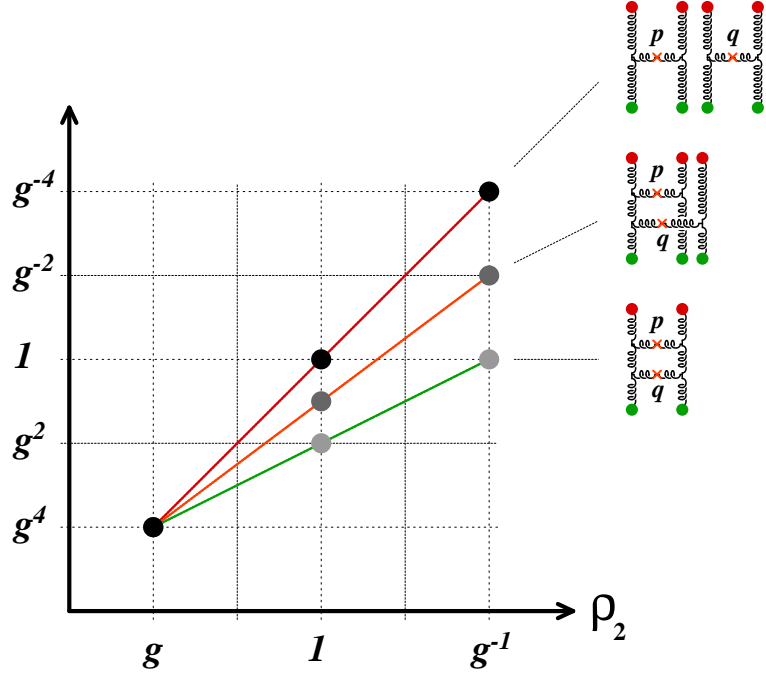}}
\caption{
Relative importance of connected and disconnected diagrams to the
two gluon correlation function. One of the color charge densities is 
considered large, $\rho_1 \sim 1/g$, whereas the  other is allowed to vary between the
``AA'' case $\rho_2 \sim 1/g$ and the ``pA'' one $\rho_2 \sim g$. The order of the
disconnected diagram, on top, is $g^4 \rho_1^4 \rho_2^4$, whereas the
 interference diagram in the middle is $g^4 \rho_1^3 \rho_2^3$ and the
 connected one, lowest, is $g^4 \rho_1^2 \rho_2^2$.
In the ``AA'' case the disconnected diagram dominates, for the ``pA'' case all three
are equally important. In the dilute ``pp'' limit only the connected diagram
matters and both gluons are produced from the same BFKL ladder.
}\label{fig:diags}
\end{figure}

In \se\ref{subsec:fact} we argued that the leading logarithmic corrections
to particle production should be factorizable into the distributions describing
the sources. As the argument is essentially one of causality, it applies
also to multigluon production (a more formal proof is given 
in Ref.\cite{Gelis:2008ad}).
The case is particularly straightforward when the
rapidity differences between the produced gluons are small enough compared
to $1/\as$, so that at leading order one does not need to resum the radiation
of additional gluons between the measured ones. In this case the rapidity structure 
is indeed a trivial one: when observing gluon fields within a rapidity window smaller
than $1/\as$; all the gluons are sensitive to the same configuration of sources
and the gluon fields are boost invariant. 

In fact, gluon correlations in AA-collisions\footnote{
By ``AA'' what is meant here is really the formal power counting situation
when the color sources of both nuclei are strong, as we shall discuss shortly.}
are in a sense simpler than in 
more dilute systems. To understand why let us 
first formally evaluate the leading contribution in a dense-dense 
collision.
Consider the probability distribution of the number of gluons
produced in a small rapidity interval.
It is convenient to define a generating functional
\begin{equation}\label{eq:moments}
\mathcal{F}[z(\mathbf{p})]=\sum_{n=0}^{\infty}\frac{1}{n!}
\int \left[
\prod_{i=1}^n
\ud^3\mathbf{p}_i\;
(z(\mathbf{p}_i)-1)\right]
\frac{\ud^n N_n}{\ud^3\mathbf{p}_1\cdots \ud^3\mathbf{p}_n}.
\end{equation}
The Taylor coefficients of $\mathcal{F}$ around $z=1$ correspond to the
moments of the probability distribution; integrated over the momenta of 
the produced gluons they are
\begin{equation}
\left\langle N \right\rangle \quad 
 \left\langle N(N-1) \right\rangle  \quad \dots 
\quad \left\langle N(N-1)\cdots(N-n+1) \right\rangle.
\end{equation}
The result of  Ref.~\cite{Gelis:2008ad} is that when these moments
are calculated to NLO accuracy, the leading logarithms 
can be resummed into the JIMWLK evolution of the sources; 
completely analogously to the single inclusive gluon distribution.
The resulting probability distribution can be written as:
\begin{equation}\label{eq:probadist}
\frac{\ud^n P_n}{\ud^3\mathbf{p}_1\cdots \ud^3\mathbf{p}_n}
=
\int\limits_{\rho_1,\rho_2}
W_{Y}\big[\rho_1\big]  W_{Y}\big[\rho_2\big]
  \frac{1}{n!}  \frac{\ud N}{\ud^3\mathbf{p}_1}
\cdots  \frac{\ud N}{\ud^3\mathbf{p}_n}
e^{ -\int \ud^3\mathbf{p} \frac{\ud N}{\ud^3\mathbf{p}}}.
\end{equation}
Here the factors $\frac{\ud N}{\ud^3\mathbf{p}_n}$ denote
the multiplicities corresponding to the leading order
classical fields.

The probability distribution~\nr{eq:probadist} has been derived to the 
leading order in a weak coupling expansion in $\as$, and resumming
leading logarithms of  $1/x$. In this power counting, each insertion 
of the external color source contributes a factor $\rho \sim 1/g$,
so that the saturation scale $g^2 \langle\rho \rho\rangle \sim \qs^2$
is taken to have no powers of the coupling. This leads to the 
glasma fields being parametrically $A_\mu \sim 1/g$ and 
$\frac{\ud N}{\ud^3\mathbf{p}_n} \sim 1/\as$ at the dominant classical level.
An additional insertion of a color source then does not change the power of 
$\as$ of the result, since the $1/g$ in $\rho$ compensates the additional $g$
in the three gluon vertex that attaches it into the diagram. Thus the leading
contribution is given by a sum of all tree diagrams with 
an arbitrary number of insertions of the classical 
source\cite{Gelis:2006yv,*Gelis:2006cr}; a computation that 
has to be done numerically.  In the language
of BFKL physics, the dominant contribution to $n$-gluon production
in this ``AA'' case comes from cutting one rung in $n$ distinct BFKL ladders.
Producing two gluons from the same ladder is suppressed by the weak coupling
compared to this ``disconnected'' contribution.
In the ``pA'' or ``pp'' power counting, on the other hand, one or both
of the sources 
$\rho$ are assumed to be weak. For example at large $x$ the color source 
consists of a number $\mathcal{O}(1)$ of charged (valence-like) partons, each
with a charge $\sim g$ (times a color factor). This leads to 
$\rho \sim g$, and additional insertions of  such a weak external
color charge lead to a contribution that is suppressed at weak coupling.
Thus in the ``pp'' case the dominant contribution to multigluon 
production is achieved by producing all the gluons from the same ladder,
minimizing the number of insertions of the color source. Whereas
in the AA-case the dominant contribution comes from (complicated) tree diagrams
and has a classical field interpretation, in  the ``pp'' case 
the $n$ gluons are produced from the same ladder, which is an $n-1$-loop diagram.
In the ``pA'' case, when $\rho_A \sim 1/g$ and $\rho_p \sim g$ both the connected
(loop) and disconnected digrams are parametrically equally important.
This structure is illustrated in 
\fig\ref{fig:diags}, where the contributions of the ``disconnected'' and 
``connected'' diagrams and their interference term are evaluated for the 
case of $\rho_1 \sim 1/g$ and $\rho_2$ having different parametric dependences
on $g$. In practice, however, many applications to two particle correlations
in the ``pA'' case only the connected part has been considered
(see e.g.~\cite{Fukushima:2008ya} for a discussion).

Note that  the
Poissonian-looking form of the result is to some extent an artifact of our choosing
to develop and truncate precisely the moments \eq\nr{eq:moments} that
are simply $\langle N \rangle^n$ for a Poissonian distribution. 
Since in our
power counting $N\sim 1/\as$, any contributions that would make the
distribution \eq\nr{eq:probadist} deviate from the functional form are
of higher order in the weak coupling expansion of the moments \nr{eq:moments} 
and are neglected in the calculation unless they are enhanced by large 
logarithms of $x$. It should also be emphasized that in spite of appearances
of \eq\nr{eq:probadist} the probability distribution is in fact not
Poissonian. To understand the nontrivial nature of this result 
it must be remembered that the individual factors
of $\frac{\ud N}{\ud^3\mathbf{p}_i}$ in \eq\nr{eq:probadist} are all functionals
of the \emph{same} color charge densities $\rho_{1,2}$; thus the averaging 
over the $\rho$'s induces a correlation between them. These correlations 
are precisely the leading logarithmic modifications to the probability distribution;
they have been resummed into the distributions $W_y$; the functional 
form of the multigluon correlation function \emph{under} the functional 
integral in \eq\nr{eq:probadist} is the same as at leading order. 
The fact that the dominant contribution is disconnected for fixed sources
and becomes correlated only when one averages over the distribution 
of $\rho$ has the physical interpretation that the domainant 
correlations in the systems are the ones enhanced by large logarithms
of $x$ and are \emph{already present in the wavefunction} before 
the collision.
Because of this the calculation of multigluon correlations is in fact 
simplified in the  strong field limit~\cite{Kharzeev:2004bw,Armesto:2006bv}.

\begin{figure}
\resizebox{\textwidth}{!}{
\includegraphics[height=5cm]{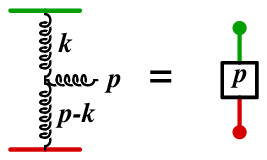}
\rule{3cm}{0pt}
\includegraphics[height=5cm]{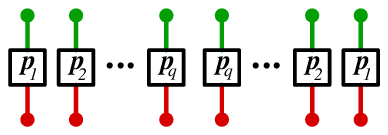}
}
\caption{Left: Building block, Lipatov vertex coupled to two sources.
Right: combinatorics of the sources. The combinatorial problem is
to connect the dots on the upper and lower side (left- and right moving 
sources) pairwise.
}\label{fig:lipatov} \label{fig:combinat}
\end{figure}

\subsection{The ridge and the negative binomial}
\label{subsec:negbin}

We can then apply this formalism to the calculation of the probability 
distribution of the number of gluons in the 
glasma~\cite{Dumitru:2008wn,Dusling:2009ar,Gelis:2009wh}.
We shall assume the ``AA'' power counting of sources that
are parametrically strong in $g$, but nevertheless work to the lowest 
nontrivial order in the color sources. Formally this would correspond
to a power counting $\rho \sim g^{\varepsilon -1}$ with a small 
$\epsilon > 0$. In this limit, as we have discussed, the dominant
contributions to multiparticle correlations come from diagrams that
are disconnected for fixed sources and become connected only
after averaging over the color charge configurations.

\begin{figure}
\centerline{\includegraphics[width=0.5\textwidth]{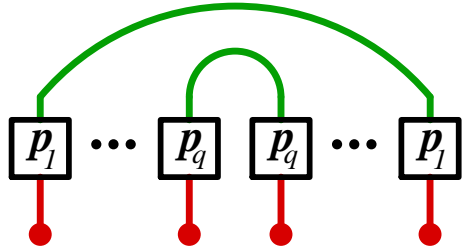}}
\caption{
Dominant contributions are ``rainbow'' diagrams, where on one
side (left- or rightmoving sources) the  same sources in the
amplitude and the complex conjugate are connected
to each other.
}\label{fig:rainbow}
\end{figure}

Working with the MV model Gaussian probability distribution
\begin{equation}
W[\rho] = \exp \left[ -\int \ud^2\xt \frac{\rho^a(\xt)\rho^a(\xt)}{g^4\mu^2} \right]
\end{equation}
computing the correlations in the linearixed approximation
 is a simple combinatorial problem. 
Each gluon is produced from two Lipatov vertices 
(see \fig\ref{fig:lipatov} left), one in the amplitude and the other in
the complex conjugate. The combinatorial factor is obtained by counting
the different ways of contracting the sources pairwise 
(see \fig\ref{fig:combinat} right).
The dominant contributions are ``rainbow'' diagrams,
\fig\ref{fig:rainbow}, where on the side of one of the sources
a line with momentum $\pt$ in the amplitude is connected to a line
with the same momentum in the complex conjugate amplitude.
These  contributions dominate because they contain a maximally 
infrared divergent integral in the momentum circulating in the lower 
(``non-rainbow'') side of the diagram. This divergence is then regulated 
by the transverse correlation scale of the problem, $\qs$.
When integrated over the momenta of the produced gluons one 
obtains the factorial moments of the multiplicity, 
which define the whole probability distribution.
The result of the combinatorial  exercise is that the number of contributing diagrams, 
each with an equal contribution, is $2^{q}(q-1)!$.
The probability distribution  can be expressed in 
terms of two parameters, the mean multiplicity $\bar{n}$, and a
parameter $k$ describing the width of the distribution.
The factorial moments
$m_q \equiv \langle N^q \rangle - \textnormal{disc.}$
are
\begin{equation}
 m_q = (q-1)! \,  k \left(\frac{\bar{n}}{k} \right)^q
\end{equation}
 with parameters $k$ and $\bar{n}$ given by
\begin{eqnarray}
k &\approx&  \frac{  (\nc^2-1)   \qs^2 S_\perp  }{2\pi}
\\
\bar{n} &=&  f_N \frac{1}{\as} \qs^2 S_\perp.
\end{eqnarray}
These moments define a \emph{negative binomial} distribution,
 which has been used 
as a phenomenological observation in high energy hadron and nuclear collisions
already for a long 
time~\cite{Arnison:1982rm,*Alner:1985zc,*Alner:1985rj,*Ansorge:1988fg,*Adler:2007fj,*Adare:2008ns}.
In terms of the glasma flux tube picture this result has a natural  interpretation.
The transverse area of a typical flux tube is $1/\qs^2$, and thus there are 
$\qs^2 S_\perp= N_{\textnormal{FT}}$  independent ones. Each of these radiates
particles independently into $\nc^2-1$ color states in a Bose-Einstein distribution
(see e.g.~\cite{Fukushima:2009er}). A sum of $k \approx N_{\textnormal{FT}} (\nc^2-1)$ 
independent Bose-Einstein-distributions is precisely what defines a negative binomial 
distribution.

\begin{figure}
\resizebox{\textwidth}{!}{
\includegraphics[height=5cm]{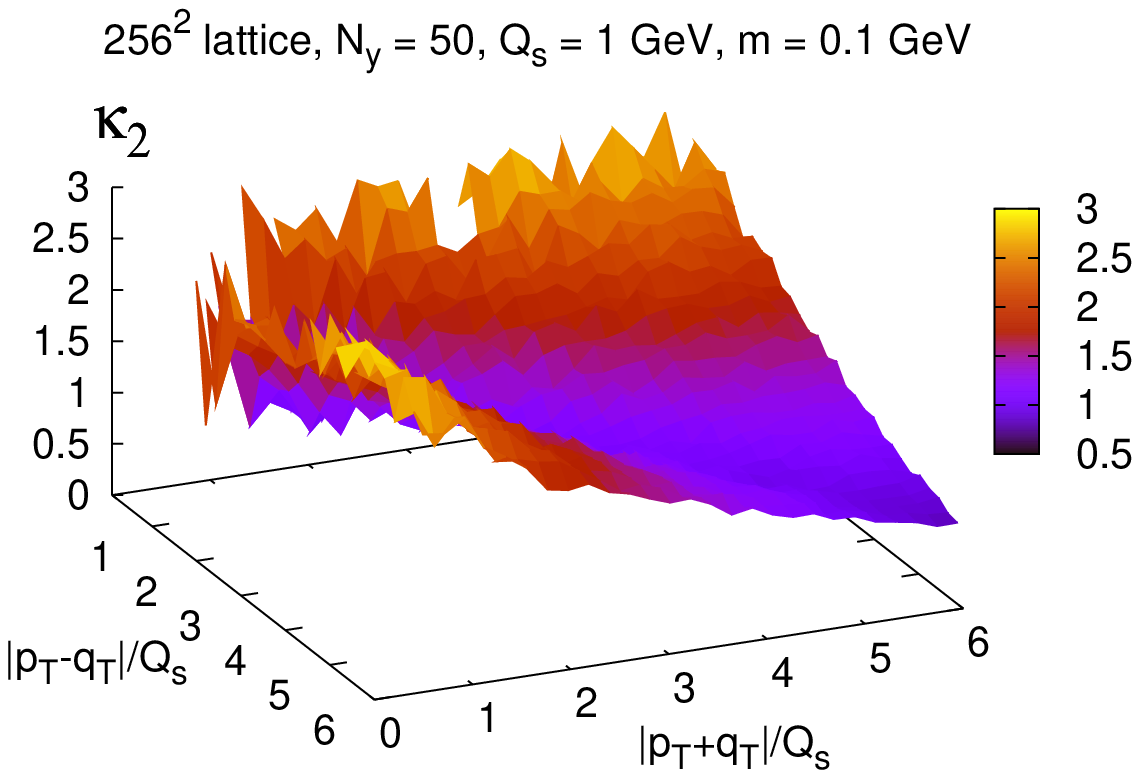}
\includegraphics[height=5cm]{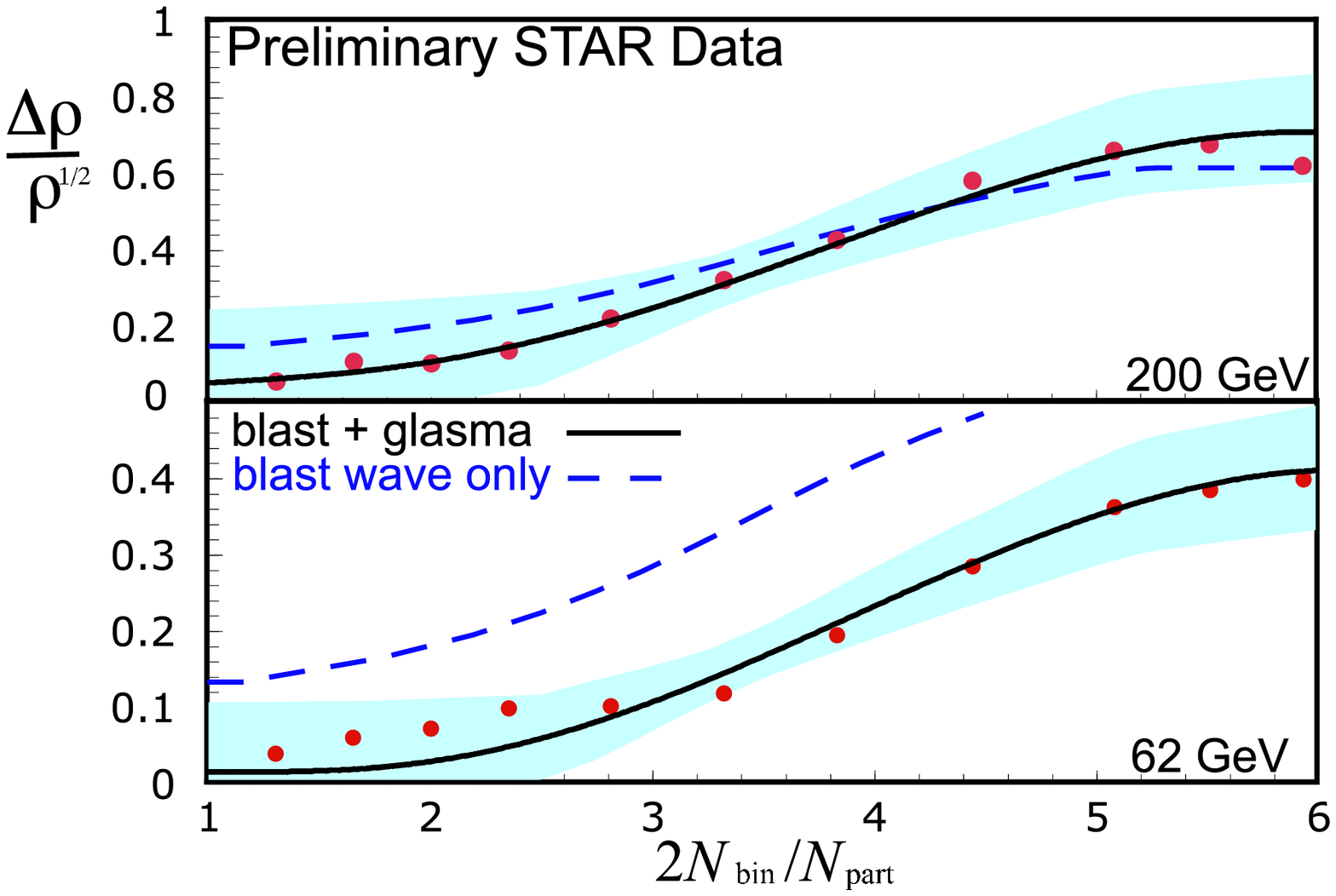}
}
\caption{
Left: Two-gluon correlation in a numerical CYM evaluation in the MV 
model\protect\cite{Lappi:2009xa}.
Right: Strength of the ridge correlation from a calculation including
the initial correlations from the glasma fields and a blast-wave parametrization
of hydrodynamical evolution\protect\cite{Gavin:2008ev}.
}\label{fig:kappa2}
\end{figure}

An important special case of this result (and one that was first obtained 
independently from the general derivation of the multiplicity distribution 
in\cite{Dumitru:2008wn}) are two gluon correlations. Similarly as in 
\se\ref{sec:bulk} one can argue thate they should be closely related to the
two particle correlations observed in the final state. There are two major 
ingredients in the ``glasma flux tube\cite{Dumitru:2008wn}'' explanation
 for the ridge:
\begin{itemize}
\item The long range longitudinal structure is provided by the approximately 
boost invariant glasma fields. The fields are correlated at a (transverse) 
length scale $1/\qs$ which enables  one to estimate the strength of the
correlation.
\item The narrow azimuthal structure is explained by a collimating 
effect\cite{Voloshin:2003ud,*Shuryak:2007fu,*Pruneau:2007ua}
from the strong radial flow generated in the hydrodynamical stage.
\end{itemize}
It has become conventional to parametrize the two particle correlation function
in terms of the single particle spectra and a geometrical factor related to the 
number of flux tubes as
\begin{equation}
\frac{C_2(\p,\q)}{
\left<\frac{\ud N}{\ud y_p \ud^2 \pt}\right>
\left<\frac{\ud N}{\ud y_q \ud^2 \qt} \right>}
=\kappa_2(\pt,\qt)\frac{1}{S_\perp \qs^2}\,, 
\label{eq:C2part2}
\end{equation}
wherer $C_2$ is the two-particle correlation function
\begin{equation}
C_2(\p,\q) \equiv \left<\frac{\ud^2 N_2}{\ud y_p \ud^2 \pt \ud y_q \ud^2 \qt }\right> 
-\left<\frac{\ud N}{\ud y_p \ud^2 \pt}\right>\left<\frac{\ud N}{\ud y_qd^2 \qt}
\right>.
\label{eq:C2part}
\end{equation}
Generally $\kappa_2$ is expected to be a number of order one. It is a constant 
$\kappa_2 \approx 0.4$\footnote{After correcting for incorrect constant factors 
in the original calculation of \protect\cite{Dumitru:2008wn}; 
see\protect\cite{Lappi:2009xa} for a discussion.}
(up to logarithms) in the dilute limit calculation\cite{Dumitru:2008wn,Gelis:2009wh},
but there is no reason for it to be a constant in a full calculation. The scaled
correlation function has recently been evaluated\cite{Lappi:2009xa} in the MV model,
with the result typically being very close to $\kappa_2 \approx 1$, but with some 
dependence on $\pt$ and $\qt$. The result of this numerical calculation 
is shown in \fig\ref{fig:kappa2} as a function of $|\pt-\qt|$ and $|\pt+\qt|$.
The experimentally observed quantity is denoted as 
$\Delta \rho/\sqrt{\rho_{\rm ref.}}(\Delta \varphi)$. Assuming a blast-wave
parametrization of the hydrodynamical evolution the value at
$\Delta \varphi=0$ can be related to $\kappa_2$ as~\cite{Dumitru:2008wn}
\begin{eqnarray}
{\Delta \rho\over \sqrt{\rho_{\rm ref.}}} \left(\Delta \varphi =0\right) &=& {dN\over dy} \cdot \frac{C_2(\p,\q)}{
\left<\frac{\ud N}{\ud y_p \ud^2 \pt}\right>
\left<\frac{\ud N}{\ud y_q \ud^2 \qt} \right>}\, (\gamma_B - \frac{1}{\gamma_B})\nonumber \\
&=& {K_N\over \as}\, (\gamma_B - \frac{1}{\gamma_B}) ,
\label{eq:flux-data1}
\end{eqnarray}
where $K_N = \kappa_2 /13.5$ for an SU(3) gauge theory 
(based on the relation between the multiplicity and $\qs^2 S_\perp$ discussed in 
\se\ref{sec:bulk}.)
Here $\gamma_B$ is the average radial boost in the framework of a blast wave model. From the 
RHIC data~\cite{Adams:2004pa,*Daugherity:2008su}, one can estimate that 
$\Delta \rho/\sqrt{\rho_{\rm ref.}}(\Delta \varphi=0)= 1/\sqrt{2\pi \sigma_\varphi^2}$, with 
$\sigma_\varphi=0.64$. Combining this with \eq\nr{eq:flux-data1}, one obtains 
\begin{equation}
\kappa_2^{\textrm{BW}} \sim {0.7 \over (\gamma_B - \frac{1}{\gamma_B})} \, ,
\end{equation}
for $\as =0.5$ and where the superscript denotes that this is a crude 
estimate extracted from experiment using a blast-wave parametrization. 
For an average blast wave radial velocity $V_r =0.6$, this gives 
$\kappa_2^{\textrm{BW}}\sim 1.5$; 
for $V_r =0.7$, one obtains $\kappa_2^{\textrm{BW}} \sim 1$.
A comparison of this parametrization to experimental data
is shown in \fig\ref{fig:kappa2}.  
For more detailed phenomenological studies we refer the reader 
to\cite{Gavin:2008ev,Moschelli:2009tg}.
In spite of the increasingly detailed modeling of the final stage effects there
remains some uncertainty in the strength of the correlation. More detailed 
understanding of the strength of the initial correlations from 
JIMWLK evolution and further comparisons of realistic hydrodynamical 
calculations are still needed to clarify these issues.

\subsection{Chiral magnetic effect}

\begin{figure}
\resizebox{\textwidth}{!}{
\includegraphics[height=5cm]{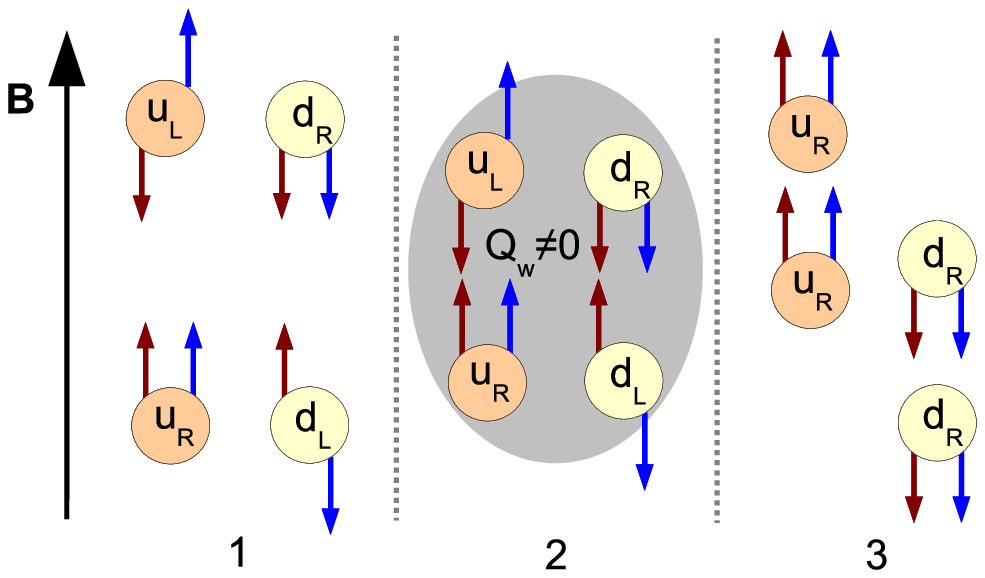}
\includegraphics[height=5cm]{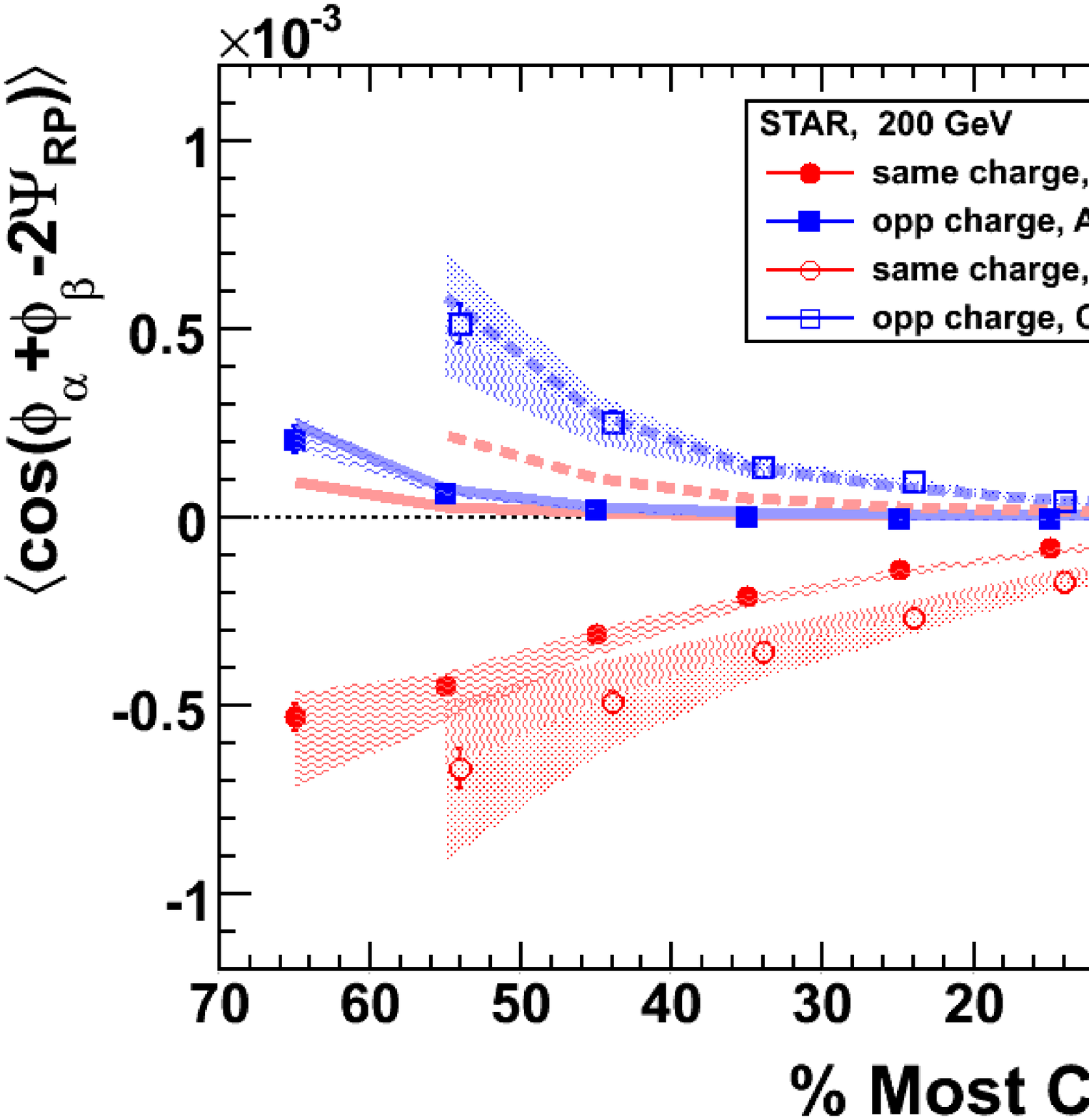}
}
\caption{
Left: Illustration of the chiral magnetic effect.
In situation 1 the spins (blue or right arrow) of the positively charged
$u$
quarks are aligned with the external magnetic field $\vec{B}$,
and those of the $d$ quarks against it. The momenta (red or left arrow)
of the right handed quarks are in the direction of the spin and 
those of the left-handed quarks are opposite to it. After an external 
color field configuration with nonzero Chern-Simons number an excess
of right-handed quarks is present. This leads to an excess of momentum
in the direction of $\vec{B}$ for the $u$ quarks and
against $\vec{B}$ for the $d$'s, i.e. an electric dipole moment.
Right: 
Correlation data showing a signal for a like side correlation 
for same charge pairs from\protect\cite{Abelev:2009uh,*Abelev:2009tx}.  
The solid (AuAu) and dashed (CuCu) lines represent the estimates of the
effect of 3-particle correlations based on HIJING.
} \label{fig:chimag}
\end{figure}

There have been several discussions\cite{Kharzeev:1998kz,*Kharzeev:1999cz,*Kharzeev:2004ey}
of the possibility of observing event-by event CP-violating fluctuations
in heavy ion collisions. The discussion originally focused
on the possibility of generating these fluctuations in the thermal 
phase of the collision process. However, the large longitudinal
classical color fields in the glasma provide a very natural framework 
for the phenomenon to occur\cite{Kharzeev:2001ev,Lappi:2006fp}.
Parity violation in a large classical field configurations can occur
when one has a large Chern-Simons charge $\ncs$, which is the time component
of the topological current $K^\mu$; 
$\partial_\mu K^{\mu} = \widetilde{F}_a^{\mu\nu}F^a_{\mu\nu}$, 
where $\widetilde{F}^{\mu\nu} = \half \epsilon^{\mu\nu\rho\sigma} F_{\rho\sigma}$.
In terms of the gauge potentials this Chern-Simons current is
\begin{equation}
K^\mu = \epsilon^{\mu \nu \rho \sigma} A^a_{\nu}
\left( F^a_{\rho\sigma} + \frac{g}{3} f^{abc}A^b_\rho A^c_\sigma \right).
\end{equation}
In the strict  boost invariant case $\ncs$ is suppressed by topologial 
reasons, which has been seen in numerical simulations\cite{Kharzeev:2001ev}.
But in the presence of fluctuations that break boost invariance the
natural (unsuppressed by the weak coupling) strength of $\ncs$ is expected
to be seen, although this has not yet been confirmed by an explicit
calculation.
When quarks are produced in the color field, a nonzero value of the 
Chern-Simons charge leads to an excess of quarks of a given chirality 
being produced. Due to symmetry the average of $\ncs$ over a large number of
events will be zero, but the fluctuations will be large and measurable.
The magnitude of the fluctuations can again be estimated using the fact that
the systems consists of correlated areas of size $1/\qs^2$ in the transverse
plane. The number of these domains,  each having charge $\sim 1$, is
$\sim\pi\ra^2\qs^2$.
The sum of these independent charges will fluctuate around zero with magnitude
$\sqrt{ \left\langle  \ncs^2 \right\rangle} 
\sim \sqrt{N} \sim \qs \ra$. 

The experimental manifestation of parity violation in heavy ion collisions
happens through the ``chiral magnetic 
effect''\cite{Kharzeev:2007jp,*Fukushima:2008xe,*Kharzeev:2009pj,*Fukushima:2009ft} 
(see \fig\ref{fig:chimag}).
In a noncentral collision there is an external magnetic field perpendicular
to the reaction plane caused by the positive charges of the ions. The spins
of the quarks become correlated with this magnetic field. In the presence
of a net chirality this causes the momenta of the quarks to be correlated
with their electric charge times the magnetic field.
This means that
one is generating, event by event,  a net electric dipole moment (a vector)
that is correlated with the reaction plane (pseudovector), a classical
signal of CP violation analogous to the elusive neutron electric dipole moment.
Such a correlation between the electrical charges and  momenta of the produced
particles with the reaction plane has now, after being predicted by the theory,
been obseved by the STAR 
experiment\cite{Voloshin:2008jx,*Voloshin:2009hr,Abelev:2009uh,*Abelev:2009tx}
(see however\cite{Wang:2009kd,*Bzdak:2009fc,*Pratt:2010gy} 
for a discussion of other effects leading to 
similar experimental signatures).

\subsection{Rapidity dependence of two gluon correlation}

\begin{figure}
\includegraphics[width=0.45\textwidth]{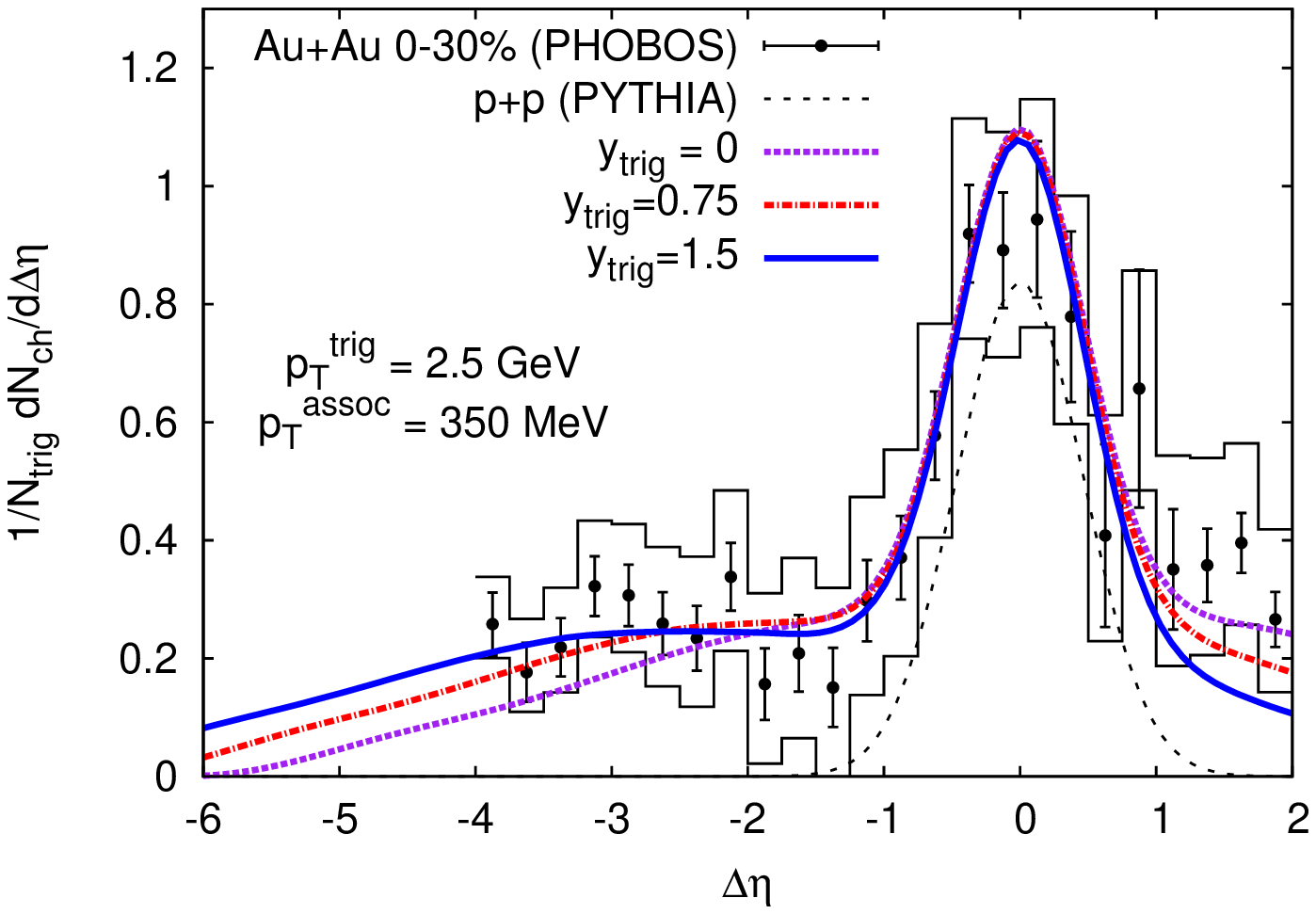}
\hfill
\includegraphics[width=0.45\textwidth]{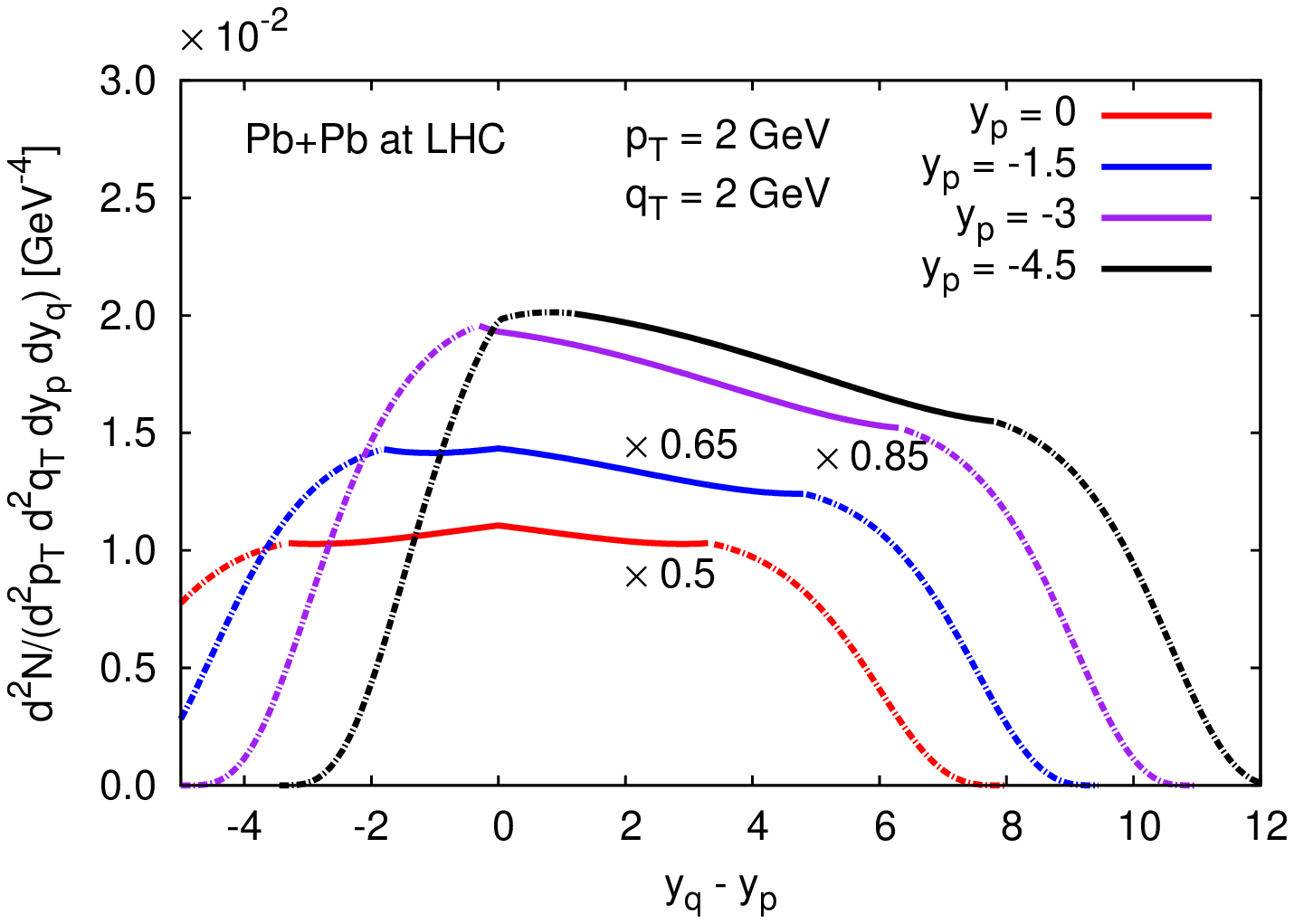}
\caption{
Left: Comparison of a two-particle correlation computed using 
\eq\nr{eq:double-inclusive-4} supplemented with a short-range correlation 
contribution from PYTHIA with PHOBOS data.
Right: comparison Left: Rapidity correlation at LHC energies $\ktt$-factorization approximation.
Plots from\protect\cite{Dusling:2009ni}.
}\label{fig:rapidity}
\end{figure}

The same power counting argument that qualitatively distinguishes
AA collisions from ones involving a dilute probe also applies to 
correlations between gluons at larger rapididy separations.
When producing two gluons from the same BFKL ladder (which is the
dominant mechanism in pp and pA) momentum conservation causes the 
dominant long range (in rapidity) correlation to be back-to-back 
in azimuthal angle. A long range correlation on the near side 
that leads to the ridge comes from diagrams that are disconnected
for fixed sources and only become connected at the level of the
source distribution. To calculate the rapidity dependence of the 
ridge one must therefore go beyond the leading order boost invariant
classical fields. The first step in this direction is to compute
the leading logarithmic corrections, more precisely to
resum corrections that are proportional to $\as(y_p-y_q)$, where
$y_p$ and $y_q$ are the rapidities of the two produced gluons. 
In the dense-dense case when the correlations are dominated by 
those present in the color sources,
this is achieved by including the JIMWLK evolution between the
rapidities of the produced gluons\cite{Gelis:2008sz,Lappi:2009fq}.

The crucial aspect of JIMWLK evolution for unequal rapidity 
correlations is to realize that the probability distribution 
$W[\rho]$ should really be understood as a distribution
for \emph{trajectories} of the color charge densities (or 
in practice the Wilson lines derived from them) along 
path to larger rapidities. In this sense the formalism
contains more physical information than just the equation for 
$W_y[U(\xt)]$; the probability distribution at a single 
rapidity $y$. 
Namely, it also gives the combined
probability distribution for Wilson lines at different rapidities
\begin{equation}\label{eq:npointdiscr}
W_{y_1 \dots y_n} [U_1(\xt), \dots, U_n(\xt)],
\end{equation}
or in a formulation in that is continuous in rapidity
$W[U(\xt,y)]$.
We can formally 
return from the distribution of trajectories to a distribution 
of Wilson lines at one individual rapidity as
\begin{equation}
W_{y}[U(\xt)]
\equiv
\int\left[DU(y,\xt)\right]\;
W\left[U(y,\xt)\right]
\delta\left[U(\xt)-U(y,\xt)\right].
\end{equation}
Knowing the general (multiple rapidity) probability distribution
will enable us to compute the correlations between Wilson lines,
and consequently of physical observables such as the multiplicities,
at different rapidities.
 A transparent interpretation of
this can be obtained in the Langevin formulation of high
energy evolution\cite{Blaizot:2002xy}, where one explicitly
constructs the ensemble of Wilson lines from solutions
of a nonlinear Langevin equation. The Langevin formulation 
is also the one used in the (numerical) full solution of the 
JIMWLK equation\cite{Rummukainen:2003ns,Kovchegov:2008mk}.

Knowing that the correlation follows from a Langevin equation 
imposes an additional structure
(of a Markovian process) on the joint probability distribution:
\begin{equation} \label{eq:markov}
W_{y_p,y_q}\left[U^p,U^q\right] 
=
G_{y_p-y_q}\left[U^p,U^q\right] 
W_{y_q}\left[U^q\right],
\end{equation}
where the JIMWLK Green's function $G$ is determined by the initial condition
\begin{equation} \label{eq:greenic}
\lim_{y_p \to y_q} G_{y_p-y_q}\left[U^p,U^q\right] 
=
\delta\left(U^p(\xt)-U^q(\xt)\right)
\end{equation}
and the requirement that it must satisfy the JIMWLK equation
\begin{equation}
\partial_{y_p} G_{y_p-y_q}\left[U^p,U^q\right] 
= \mathcal{H}\left(U^p(\xt)\right)G_{y_p-y_q}\left[U^p,U^q\right].
\end{equation}
This JIMWLK Green's function contains all the information, at the leading 
log level, of long range rapidity correlations in gluon production. 
This structure follows from the
computation of the leading log part of 1-loop corrections  to  a 
wide class of observables that can be expressed in terms of correlators
of the gluon fields at $\tau=0$ (or equivalently in terms of the Wilson lines) 
\begin{equation}
\left<{\cal O}\right>_{_{\rm LLog}}
=
\int 
\left[DU_1(y,\xt)\right]\left[DU_2(y,\xt)\right]
W\left[U_1(y,\xt)\right]W\left[U_2(y,\xt)\right]
\mathcal{O}_{_{\rm LO}}.
\label{eq:fact-gen}
\end{equation}

A full (numerical) calculation of rapidity correlations in
JIMWLK evolution is yet to be done. The first
phenomenological applications of this framework\cite{Dusling:2009ni}
follow the strategy of using the approximate linearized solution 
of the equations of motion used also in the boost invariant 
case reviewed in \se\ref{subsec:negbin}.  One solves the 
two gluon correlation in terms of $\rho \rho$-correlators\footnote{
We are here assuming
a Gaussian distribution so that no nontrivial higher cumulants
contribute.}, including unequal rapidity ones.
Recognizing that due to the Markovian nature of the evolution
$\left\langle\rho_y^{A}(\xt)\rho_{y'}^{A}(\yt) \right\rangle
=
\left\langle\rho_y^{A}(\xt)\rho_{y}^{A}(\yt) \right\rangle$,
where $y$ is the rapidity that is earlier in the evolution of
nucleus A than $y'$, one can express the result in terms
of equal-rapidity correlations of two $\rho$'s. To get a
realistic  estimate of
the resulting rapidity dependence one then replaces the
equal rapidity $\rho \rho$ correlators by solutions of the 
(running coupling) BK equation, denoted below as
$\Phi(y,\kt)$. The resulting $\ktt$-factorized approximation is
\begin{align}\label{eq:double-inclusive-4}
C_2(\p,\q)
& =
\frac{\alpha_s^{2}}{16 \pi^{10}}
\frac{N_c^2(N_c^2-1) S_\perp}{d_A^4\; \pt^2\qt^2}
\int \ud^2\kt\;
\\ \nonumber
&
\bigg\{
\Phi_{A_1}^2(y_p,\kt)\Phi_{A_2}(y_p,\pt-\kt)
\Big[
\Phi_{A_2}(y_q,\qt+\kt)
+
\Phi_{A_2}(y_q,\qt-\kt)
\Big]
\\ \nonumber
&
+
\Phi_{A_2}^2(y_q, \kt)\Phi_{A_1}(y_p,\pt-\kt)
\Big[
\Phi_{A_1}(y_q,\qt+\kt)
+
\Phi_{A_1}(y_q,\qt-\kt)
\Big]
\bigg\}\, .
\end{align}
Although the required symmetrizations make this formula look a bit awkward,
the structure is in fact quite simple. The double inclusive
gluon multiplicity is proportional to four unintegrated distributions
(the single inclusive one being given by a convolution 
of two). This is a demonstration of our power counting argument; when 
the distributions are assumed to be large $\Phi \sim 1/\as$, this contribution 
to the correlation is $\sim 1/\as^2$. Producing the gluons from the 
same one BFKL ladder (``pp'' contribution) would be parametrically
$\as^2 \Phi^2 \sim 1$. A result of the Markovian nature of the correlation 
is that the rapidity arguments are not symmetric; for example in the first 
term three out of the four distributions are evaluated at the rapidity 
$y_p$ (earlier in the evolution of nucleus 1) and only one at the rapidity 
$y_q$.
Implementing \eq\nr{eq:double-inclusive-4} leads to a very weak 
rapidity dependence and a normalization of the correlation in broad agreement 
with the experimental results, as seen in \fig\ref{fig:rapidity}. One 
could assume that the decorrelation with rapidity could be faster
in the full nonlinear JIMWLK evolution than the linearized version 
assumed in deriving the $\ktt$-factorized approximation
of the correlation. This would require combining the numerical 
JIMWLK evolution with a full CYM calculation of the multiplicity, 
which has not yet been performed.

\section{Conclusions}

We have emplasized  in this short review the potential of the CGC 
framework as a description of the initial 
conditions of a heavy ion collision, based on weak coupling 
QCD. The CGC picture enables one to analyze DIS at
small $x$, proton-nucleus collisions and the initial conditions
in AA collisions in a single unified framework. As a weak-coupling
first-principles QCD calculation there is a systematic way to
compute higher order corrections to the results. We have seen that the 
basic features of bulk particle production in AA collisions, such as the 
collision energy, rapidity and centrality dependence can be well understood
in terms of a single dominant momentum scale $\qs$ present in the 
high energy nuclear wavefunction.
An area of recent and ongoing 
activity are multiparticle correlations, which have a high 
potential of giving direct experimental access to initial state
of the collision.

\section*{Acknowledgements}
The author is supported by the Academy of Finland, project 126604.

\bibliographystyle{JHEP-2modM}

\bibliography{spires}

\end{document}